\documentclass[useAMS,usenatbib]{mn2e}
\usepackage{graphicx}

\title[PGMS: observations and CMB foreground analysis]
  {The Parkes Galactic Meridian Survey (PGMS): observations and
       CMB polarization foreground analysis}

\author[E. Carretti et al.]
  {E.~Carretti,$^{1,2}$\thanks{E-mail: Ettore.Carretti@csiro.au (EC)}
  M.~Haverkorn,$^{3,4,5}$ D.~McConnell,$^6$ G.~Bernardi,$^7$
  \newauthor 
  N.M.~McClure-Griffiths,$^6$ S. Cortiglioni,$^8$ and S.~Poppi,$^9$\\
  $^1$ATNF, CSIRO Astronomy and Space Science, P.O. Box 276, Parkes, NSW 2870, Australia\\
  $^2$INAF, Istituto di Radioastronomia, Via Gobetti 101, I--40129 Bologna, Italy\\
  $^3$Jansky Fellow, National Radio Astronomy Observatory\\
  $^4$Astronomy Department, University of California, Berkeley, 601 Campbell Hall, Berkeley, CA 94720\\
  $^5$ASTRON, Oude Hoogeveensedijk 4, 7991 PD Dwingeloo, The Netherlands\\
  $^6$ATNF, CSIRO Astronomy and Space Science, P.O. Box 76, Epping, NSW 1710, Australia\\
  $^7$Kapteyn Astronomical Institute, University of Groningen, P.O. Box 800, 9700 AV Groningen, the Netherlands\\
  $^8$INAF, Istituto di Astrofisica Spaziale e Fisica Cosmica Bologna, Via Gobetti 101, 40129 Bologna, Italy\\ 
  $^9$INAF, Osservatorio Astronomico di Cagliari, Loc. Poggio dei Pini, Strada 54, 09012 Capoterra, Italy
  }

\begin{document}

\date{Accepted xxxx. Received yyyy; in original form zzzz}

\pagerange{\pageref{firstpage}--\pageref{lastpage}} \pubyear{2010}

\maketitle

\label{firstpage}

\begin{abstract}
We present observations and CMB foreground analysis of the Parkes Galactic 
Meridian Survey (PGMS), an investigation of the Galactic latitude behaviour of
the polarized synchrotron emission at 2.3 GHz with the Parkes Radio  
Telescope. The survey consists of a $5\degr$ wide strip along the Galactic  
meridian $l=254\degr$ extending from Galactic plane to South Galactic pole.
We identify three zones distinguished by polarized emission  
properties: the disc, the halo, and a transition region connecting them. The halo section lies at  
latitudes $|b| > 40\degr$ and has weak and smooth polarized
emission mostly at large scale with steep angular power spectra of median
slope $\beta_{\rm med} \sim -2.6$.
The disc region covers the latitudes $|b|<20\degr$ and has a brighter,
more complex emission dominated by the small scales with flatter spectra of median
slope $\beta_{\rm med} = -1.8$.  The transition region has steep spectra as in  
the halo, but the emission increases toward the Galactic plane from halo  
to disc levels. The change of slope and emission structure at $b \sim -20\degr$ is  
sudden, indicating  a sharp disc-halo transition.
The whole halo section is just one environment extended over $50\degr$ with very 
low emission which, once scaled to 70~GHz, is equivalent to the CMB $B$--Mode emission
for a tensor--to--scalar perturbation power ratio $r_{\rm halo} = (3.3\pm0.4)\times10^{-3}$.  
Applying a conservative cleaning procedure, we estimate an $r$ detection limit
of $\delta r \sim 2\times 10^{-3}$ at 70~GHz (3-sigma C.L.) and,
assuming a dust polariztion fraction $<$~12\%, $\delta r \sim 1\times
10^{-2}$ at 150~GHz.  The 150~GHz limit matches the goals of planned sub-orbital experiments,
which can therefore be conducted at this high frequency. 
The 70~GHz limit is close to the goal of proposed next generation space missions, 
which thus might not strictly require space-based platforms.
\end{abstract}

\begin{keywords}
Cosmology: CMB -- Galaxy: disk -- Galaxy: halo --  polarization
\end{keywords}

\vskip 3cm

\section{Introduction}
\label{intro:Sect}

The study of the Galactic polarized synchrotron emission is essential for 
two cutting-edge fields of current astrophysics research:
the detection of the $B$--Mode of the Cosmic Microwave Background (CMB), 
for which the Galactic synchrotron is foreground emission, and 
the investigation of the magnetic field of the Galaxy.

The CMB $B$--mode is a direct signature of the
primordial gravitational wave background (GWB) left by inflation
(e.g., \citealt*{Kamionkowski98,boyle06}). 
The amplitude of its  angular power spectrum is proportional to the GWB
power, which is conveniently expressed relative to the amplitude of
density fluctuations---the so-called ``tensor-to-scalar perturbation power ratio''
$r$. \footnote{ 
Refer to \citet{peiris03} Eq.~10 for a full definition of $r$.
}
Still undetected, the current upper limit is set to $r < 0.20$ (95\% C.L., \citealt{komatsu09}) 
by the results of the Wilkinson Microwave Anisotropy Probe (WMAP).
A detection of the $B$--Mode would be evidence for primordial
gravitational waves, and a measurement of $r$ 
would help distinguish among several inflation models 
and investigate the physics of the early stages of the Universe.

Reaching this spectacular science goal will be difficult because of
the tiny size of the CMB $B$-Mode signal,
fainter than the current upper limit of  0.1~$\mu$K, and perhaps as
faint as the  1~nK corresponding to the smallest $r$ accessible by CMB ($r\sim10^{-5}$, \citealt*{amarie05}).
At such low levels the cosmic signal is easily obscured
by the Galactic foreground of the synchrotron and dust emissions.

\begin{figure}
\centering
  \includegraphics[angle=00, width=1.0\hsize]{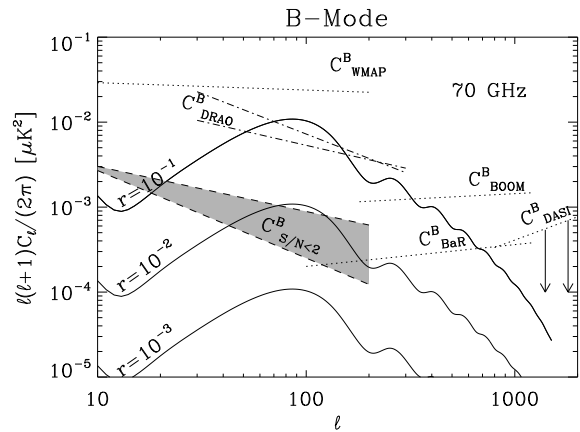}
\caption{        
         Summary of the current knowledge of the $B$--Mode power spectra of the Galactic synchrotron emission.
         Spectra are estimated at 70~GHz.  
         A brightness temperature frequency spectral index of $\alpha = -3.1$
         has been assumed for all extrapolations.   
         $C^B_{\rm WMAP}$ is the general contamination at high Galactic latitude,
         as estimated by the WMAP team using a $\sim75\%$ sky fraction at 22.8-GHz 
         \citep{page07}.
         Spectra measured in small areas selected for 
         their low emission are also shown: the target fields of 
         the experiments BOOMERanG ($C^B_{\rm BOOM}$, \citealt{carretti05b}) 
         and BaR-SPOrt ($C^B_{\rm BaR}$, \citealt{carretti06a}), 
         and the upper limit found in the two fields of the experiment DASI ($C^B_{\rm DASI}$, 
         \citealt{bernardi06}).
         Finally, $C_{S/N < 2}$ shows
         the estimate of the emission in the best 15\% of the sky (\citealt{carretti06b},
         the shaded area indicates the uncertainties), while $C_{\rm DRAO}$ the values found
         in some intermediate regions using 1.4~GHz data \citep{laporta06}. 
         For comparison, CMB spectra for three values of $r$ are also shown.
\label{status:Fig}
}
\end{figure}

Investigations of the synchrotron contribution have been conducted
over recent years, 
but data are still insufficient to give a comprehensive view
(Fig.~\ref{status:Fig}). 
\citet{page07} analysed the 23~GHz WMAP polarized maps and
find that the typical emission at high Galactic latitude is strong: at 70~GHz
it is {\it equivalent to}\footnote{
 By {\it equivalent to $r$} we signify the strength of a foreground whose spectrum 
would match the spectrum of the CMB $B$-mode emission at the $\ell=\sim90$ peak 
arising from conditions characterised by the given value of $r$.
}
$r \sim 0.3$, even higher than the current upper limit. 
An analysis of the same WMAP data by \citet{carretti06b} identified
regions covering about 15 per cent of the sky with much lower emission
levels, offering a better chance for $B$--Mode detection.  
In these regions the polarized foreground is fainter, 
equivalent to a $B$--Mode signal corresponding to $r$ in the range $[1\times10^{-3},
1\times10^{-2}]$.  A better characerization of the polarized
foreground is crucial especially for sub-orbital 
experiments  (ground-based and balloon-borne), 
which will observe small sky areas.

The design of future experiments is dependent on
the frequency of minimum foreground emission. 
WMAP finds that this is in the range 60--70~GHz for high Galactic latitudes \citep{page07},
but it remains unknown for the lowest emission part of the sky. 
It has been suggested that the dust emission might have deeper minima than the synchrotron 
in the areas of lowest emission (e.g.,~\citealt{lange08}), shifting
the best window for $B$--Mode detection 
to higher frequencies.

Synchrotron emission from the Milky Way is not only a foreground for
CMB polarization measurements, but can also be used to study the
Galactic magnetic field. The total intensity of synchrotron emission
can be used to estimate the total magnetic field strength, while the
polarized intensity gives the strength of the regular component.
This analysis in external galaxies has shown that the
spiral arms are usually dominated by a small-scale, tangled, magnetic
field with a weaker coherent large-scale field aligned with the
arms. In the inter-arm regions the regular component dominates and in
some spirals, magnetic arms with coherent scales up to the size of the
disc have been detected {\it in between} the gas arms (e.g.\ see Beck
2008 for a review).

The synchrotron emissivity of our own Galaxy is harder to understand
because of our location inside it, but has the advantage that it can
be studied in detail. 
Frequency dependent synchrotron depolarization can be used to
determine typical scale and strength of small-scale magnetic fields
(e.g.\citealt{gaensler01}), and all-sky synchrotron emissivity maps
can characterise the synchrotron scale height 
\citep{beuermann85}, or can be used for large-scale modeling of the Galactic
magnetic field, especially in the 
halo\footnote{Galactic halo is used
in this context as the gaseous and magnetic field distributions out of
the Galactic disc, and is not necessarily connected to the stellar halo.}. 
The relative parity of the toroidal magnetic field component is still under
discussion (see e.g., \citealt{han97,frick01,sun08}). 
\citet{jansson09} use WMAP synchrotron maps at 23~GHz to show 
that the magnetic field behaviour in the Galactic disc and halo may
differ considerably.

Data from external galaxies does not help in constraining the Milky
Way magnetic halo, as there is a wide variety of magnetic field
configurations: from galaxies without evident halo field, to X-shaped
fields centred at the galaxy centre, to large almost spherical
magnetic halos (see Beck 2008 for a review).

Recent maps of polarized Galactic synchrotron radiation at 1.4~and
22.8~GHz \citep{wolleben06,testori08,page07}
show polarized emission across the entire sky, and can be used
to study the Galactic magnetic field. However, the 1.4~GHz maps show
that the disc emission is strongly depolarized up to latitudes $|b|
\approx 30^{\circ}$ \citep{wolleben06,testori08},
while Faraday depolarization effects are still present up to $|b|
\approx 40 - 50^{\circ}$  \citep{carretti05a}. Furthermore, those
data consist of a single frequency band and do not enable rotation
measure computations. The WMAP data at 22.8~GHz are virtually
unaffected by Faraday rotation (FR) effects, but the sensitivity is not
sufficient since, once binned in $2\degr$ pixels, about 55\% of the sky has a
signal to noise ratio $S/N < 3$. This area corresponds to all the high
Galactic latitudes with the exception of large local structures, which
is most of the sky useful both for CMB studies and to investigate the
Galactic magnetic field.  

Therefore, synchrotron maps at intermediate
frequencies over all Galactic latitudes are needed to explore the
behaviour of the contamination of the CMB with latitude as well as to
study the Galactic magnetic field in the disc, the halo, and the
disc-halo transition.

In this work we  present the Parkes Galactic Meridian Survey (PGMS), a  
survey conducted with the Parkes Radio Telescope to cover a strip along an  
entire southern Galactic meridian at 2.3~GHz. The area is free from large local  
structures, making it ideal for investigating 
both the CMB foregrounds and the Galactic magnetic field.
The PGMS overlaps the target area of several CMB experiments like
BOOMERanG \citep{masi06}, QUaD \citep{brown09}, BICEP  
\citep{chiang09}, and EBEX  \citep{grainger08}.
Our results may have direct implications for all these experiments.

In this paper we present the survey, observations, and a characterisation
of the polarized emission. We also present an analysis of the measured emission
as a  contaminating foreground to CMB $B$--mode studies.
Analysis and implications for the Galactic magnetic field
will be subject of a forthcoming paper (Paper II, Haverkorn et al.  
2010 in preparation). A third paper will deal with the polarized extragalactic sources  
(paper III, Bernardi et al. 2010 in preparation).

Survey and observations are presented in Section~\ref{pgms:Sect}, 
the ground emission analysis in Section~\ref{ground:Sect}, and the maps 
in Section~\ref{maps:Sect}. 
The analysis of both the angular power spectrum and emission behaviour 
is presented in Section~\ref{aps:Sect}, while the dust contribution is 
investigated in Section~\ref{dust:Sect}. 
The detection limits of $r$ are discussed in Section~\ref{limits:Sect} and, finally, 
our summary and conclusions are reported in Section~\ref{conc:Sect}.

\section{The Parkes Galactic Meridian Survey}
\label{pgms:Sect}

The available data and the properties of the synchrotron emission discussed in Section~\ref{intro:Sect}
lead to the following main requirements for a survey. Observations must
\begin{enumerate}  
  \item{} be conducted at a low enough radio frequency for the synchrotron emission to dominate
               the other diffuse emission components, but at a frequency higher than 1.4~GHz to avoid
               significant Faraday Rotation effects;
  \item{}  cover all latitudes from the Galactic plane to the pole, to explore the 
               behaviour with the Galactic latitude $b$;
  \item{}  cover regions free from large local structures, such as the big radio loops, that would 
               distort the estimates of typical conditions at high latitudes.
\end{enumerate}  

\begin{figure*}
\centering
  \includegraphics[angle=90, width=1.0\hsize]{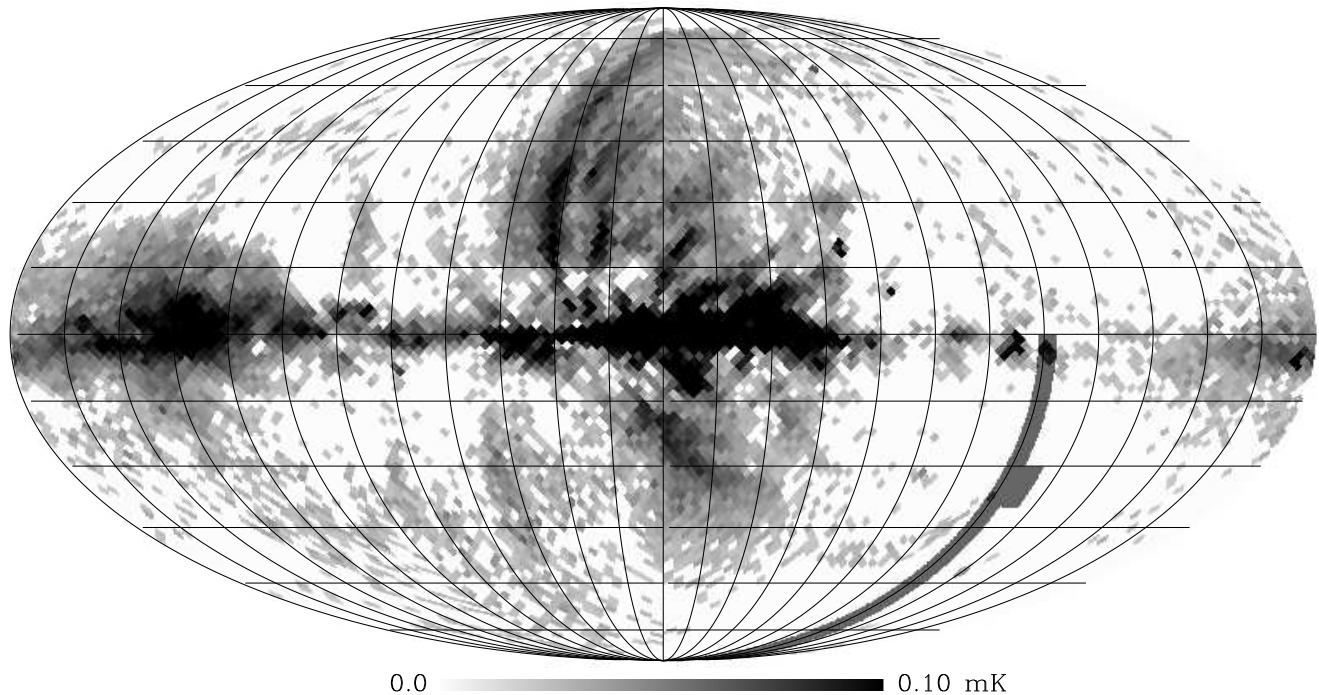}
\caption{  The PGMS strip (dark gray) plotted on the WMAP polarized intensity map ($L=\sqrt{Q^2+U^2}$)
                  at 22.8 GHz \citep{page07}
                  binned in $2^\circ$ pixels (HEALPIX pixelation with Nside=32). 
                  Pixels with S/N $<$ 3 have been blanked (white). The PGMS goes
                  through a region clear of large emission structures avoiding contamination by
                  major large local anomalies, such as the big radio loops and the fan region (centred at $l\sim135^\circ$,  $b\sim0^\circ$).
                  The big radio loops are structures large several tens of degrees, like loop-I which extends from
                  the north Galactic pole down to the Galactic plane, with a possible continuation into 
                  the southern hemisphere.
                  The map is in Galactic coordinates with longitude $l=0^\circ$ at centre and increasing leftward. 
\label{cover:Fig}
}
\end{figure*}

The Parkes Galactic Meridian Survey (PGMS) is a project to survey the diffuse polarized emission 
along a Galactic meridian designed to satisfy these requirements. 
It surveys a $5\degr\times90\degr$ strip along the entire southern 
meridian $l=254^\circ$ from the Galactic plane to the south Galactic pole (Fig.~\ref{cover:Fig}). 
The observations have been made at 2.3~GHz with the Parkes Radio Telescope (NSW, Australia),
a facility operated by the ATNF - CSIRO Astronomy and Space Science\footnote{http://www.atnf.csiro.au}
a division of CSIRO\footnote{http://www.csiro.au}.
It also includes an $10\degr \times 10\degr$ extension centred at $l=251^\circ$ and
$b=-35^\circ$.

The selected meridian goes through one of the low emission regions
of the sky identified using the WMAP data (Fig.~\ref{cover:Fig}, 
see also~\citealt{carretti06b}) and is free of large local emission structures.
The meridian also goes through the area of deep polarization observations of
the BOOMERanG experiment~\citep{masi06}; 
the {10\degr} extension near $b=-35^\circ$ is positioned to best
cover that field.

At long wavelengths, measurements of Galactic polarized emission in regions 
of high rotation measure are corrupted by Faraday depolarization. At 1.4~GHz 
Faraday depolarization is significant up to Galactic latitudes 
$|b|<40$--$50\degr$ \citep{carretti05a} where $RM > 20$~rad/m$^2$. 
At 2.3~GHz this $RM$ limit increases to $60$~rad/m$^2$, allowing a 
clear view of polarized emission over all high Galactic latitudes and well 
into the upper part of the disc.

The observations were made in four sessions from January\ 2006 to
September\ 2007 with the 
Parkes S-band Galileo receiver, named after NASA's Jupiter exploration
probe for which the Parkes telescope and this receiver were used
for down-link support \citep{thomas97}.  
The receiver responds to left- and right-handed
circular polarization, whose cross-correlation gives Stokes parameters 
$Q$ and $U$ (e.g. \citealt{kraus86}). This scheme provides more protection
against total-power (gain) fluctuations than the alternative: a receiver
responding directly to the linearly polarized signals.

The original feed used for the Galileo mission has been replaced by a
wide-band corrugated horn, highly tapered to reduce sidelobes and the
response to ground emission. 
The feed illuminates the dish with a 20~dB edge taper and the first side-lobe is
30~dB below the main beam. 

The ATNF's Digital Filter Bank~1
(DFB1) was used to produce all four Stokes parameters, $I$, $Q$, $U$,
and $V$.  DFB1 was equipped with an 8-bit ADC and  configured to
give 256~MHz spectrum with 128 2-MHz channels.  
Spectra are generated using polyphase filters that provided high 
spectral channel isolation. 
The isolation between adjacent channels is 72~dB, an enormous
improvement over the 13~dB isolation of Fourier-based correlators.
This, in combination with the high sample precision, gives excellent
protection against RFI leaking from its intrinsic frequency to other
parts of the measured spectrum. This is valuable in the 13-cm band as
strong RFI can be present.\footnote{The first observing session, in
September 2005, used a Fourier-based correlator, and spectra were
strongly contaminated by RFI.  Subsequent use of DFB1 greatly improved
the measurements.}  Recording spectra with spectral resolution greater
than required for the polarimetry analysis has allowed efficient
removal of RFI-effected channels, maximising the effective useful
bandwidth.  Data were reduced to 30 8-MHz channels.
The RFI removal typically yielded an effective total bandwidth of 160~MHz.

The source B1934-638 was used for flux calibration assuming the polynomial model by 
\citet{reynolds94} 
for an accuracy of 5\%.
The polarization response was calibrated using the sources 3C~138 and PKS~0637-752, 
whose polarization states were determined using the Australia Telescope Compact Array (ATCA) 
with an absolute error of $1^\circ$. 
The statistical error of our polarization angle calibration is $0.5^\circ$.
The astronomical IAU convention for polarization angles is used:  angles are measured from the
local northern meridian, increasing towards the east. 
It is worth noting that this differs from the convention used in the WMAP data, for which 
the polarization angle increases westwards. 
The unpolarized source B1934-638 was also used to measure the polarization leakage,
for which we measure a value of 0.4\%. The off-axis instrumental polarization
due to the optics response is about 1\%.

\begin{figure}
\centering
  \includegraphics[angle=90, width=1.0\hsize]{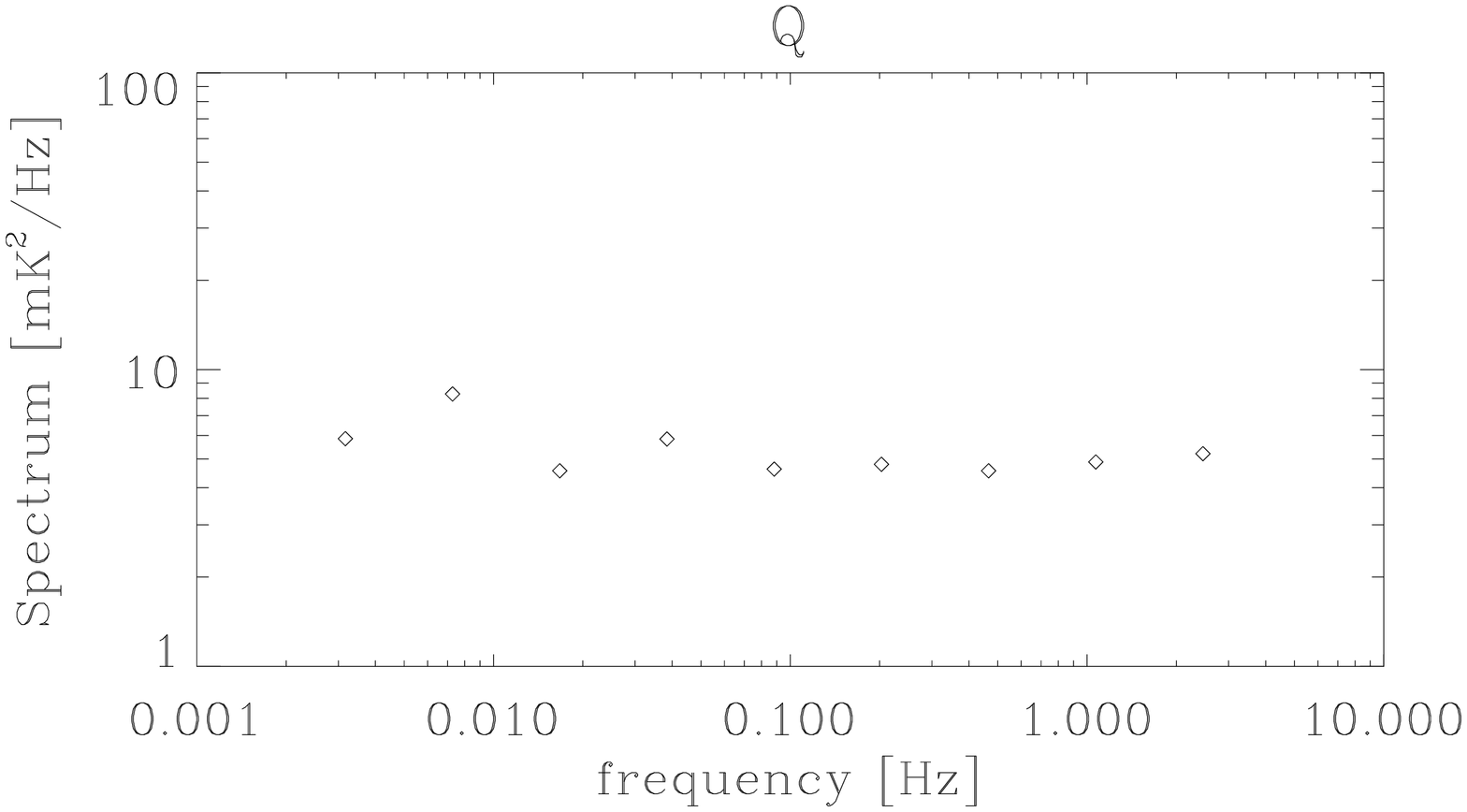}
  \includegraphics[angle=90, width=1.0\hsize]{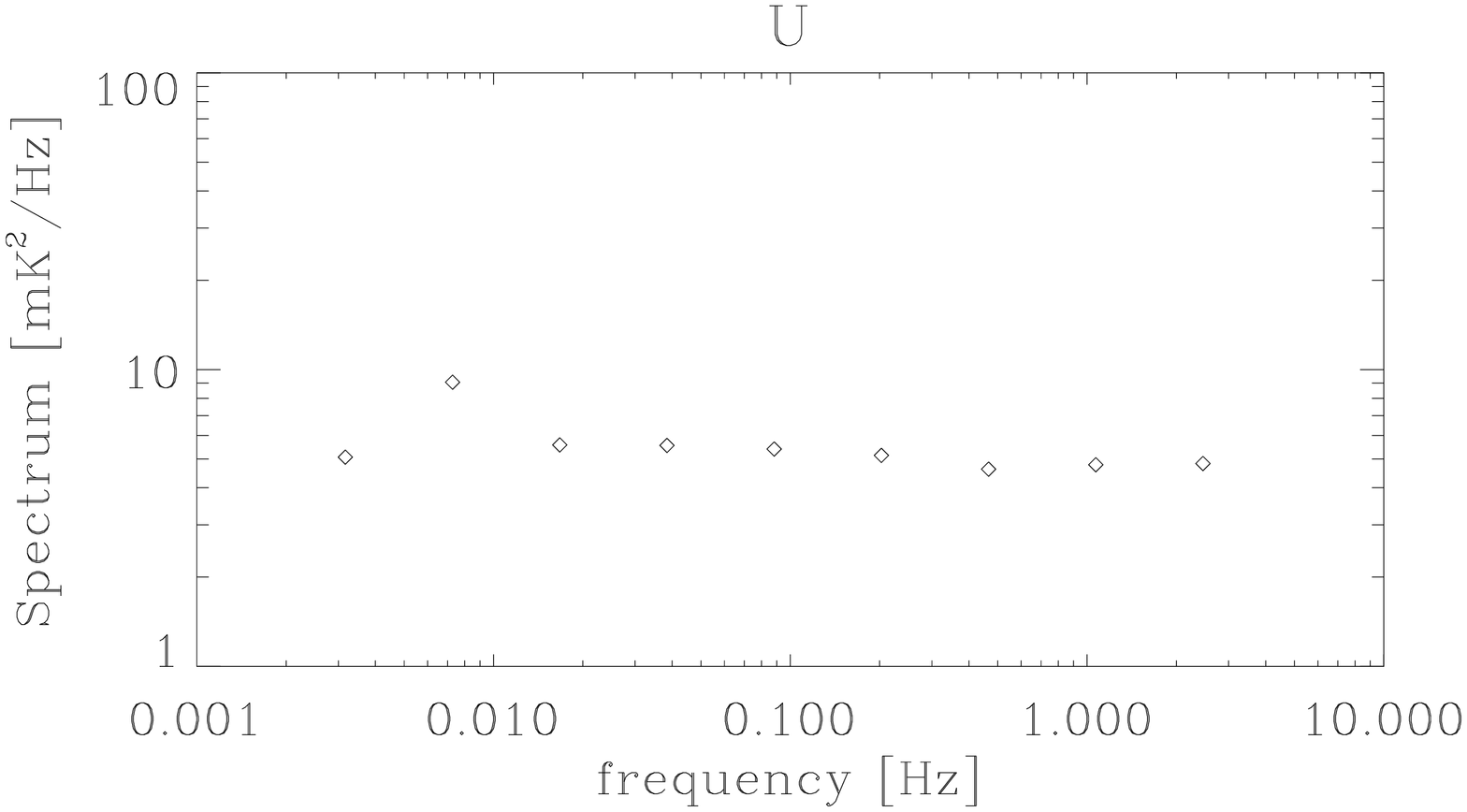}
\caption{Power spectra of $Q$ (top) and $U$ (bottom) 
time series for a south Celestical pole observation. Both spectra are mostly 
flat with no evidence of a 1/f component up to 3~mHz. 
\label{one_over_f:Fig}
}
\end{figure}

The use of a system with both $Q$ and $U$ as correlated outputs 
mitigates gain fluctuation effects. 
To check the level of a 1/f noise component in the data we observed
the South Celestial Pole (SCP), thereby avoiding azimuth (AZ) and elevation (EL)
dependent variations of atmospheric and ground emissions. 
Remaining variations in the signal arise from intrinsic atmospheric changes
and receiver fluctuations.  Power spectra of the $Q$ and $U$ time-series
are almost flat with no evidence of a 1/f component down to 3~mHz 
(see Fig.~\ref{one_over_f:Fig}). This confirms that the system is stable and 
characterised by white noise up to 7-min time scales, sufficient for the
duration of our scans.

The Galactic meridian was observed in 16 $5^\circ\times5^\circ$ fields 
and one $10^\circ\times10^\circ$ field. 
The fields are named PGMS-XX, where XX is the Galactic latitude of the field centre. 
Each field except PGMS-02 includes a $1^\circ$ extension along $b$ at the north edge 
for an actual size of  $5^\circ\times6^\circ$  and an overlap of $5^\circ\times1^\circ$ 
with the next northern field.

The fields were observed with sets of orthogonal scans to give $l$--
and $b$--maps (scans along Galactic longitude and latitude,
respectively).  Each field was observed with 101 latitude scans ($b$--maps) and 121
longitude scans ($l$--maps) spaced by 3~arcmin to ensure full Nyquist sampling
of the beam (FWHM~=~8.9~arcmin).  The same sample spacing was
used along each scan by scanning the telescope at $3^\circ$/min
with a 1-second integration time.

One full set of $l$-- and $b$--maps were observed for the 6~disc fields at 
latitude $|b| < 30^\circ$ and the $10^\circ\times10^\circ$ field, 
giving final a sensitivity of $\sim 0.5$~mK per beam-sized pixel.
Over the ten high latitude fields ($|b| > 40^\circ$), where a weaker signal was expected,
two full passes were made to give a  sensitivity of  $\sim 0.3$~mK per beam-sized pixel.

Prior to map-making, a linear baseline fit was removed from each scan and the ground emission 
contribution was estimated and cleaned up by the procedure described in Section~\ref{ground:Sect}.

The map-making procedure is based on the algorithm by Emerson \& Gr\"ave (1988), 
which combines  $l$-- and $b$--maps in Fourier space and recovers the power along 
the direction orthogonal to the scan, otherwise lost through the baseline removal. 
The algorithm is highly efficient and effectively removes  residual stripes.

Table~\ref{feat:Tab} summarises the main features of the PGMS observations.

\begin{table}
 \centering 
  \caption{Main features of the PGMS observations.}
  \begin{tabular}{@{}lc@{}}
  \hline
  Central frequency                   &  2300~MHz \\
  Effective bandwidth                & 240~MHz \\
  Useful bandwidth$^1$          & 160~MHz \\
  FWHM                                      & $8.9$~arcmin \\
  Channel bandwidth               & 8~MHz \\
  Central Meridian                    & $l_0 = 254^\circ$ \\
  Latitude coverage                   & $b = [-90^\circ, 0^\circ]$ \\
  Area size                                 & $5^{\circ} \times 90^{\circ}$ \\
  Pixel size                                & $3\arcmin \times 3\arcmin$ \\
  Observation runs                  & Jan 2006\\
                                                   & Sep 2006 \\
                                                   & Jan 2007 \\
                                                   & Sep 2007 \\
  $Q$, $U$ beam-size pixel rms sensitivity (halo fields) & 0.3~mK \\
  $Q$, $U$ beam-size pixel rms sensitivity (disc fields) & 0.5~mK \\
  \hline
  $^1$~After RFI channel flagging.
  \end{tabular}
 \label{feat:Tab}
\end{table}

\section{Ground emission}
\label{ground:Sect}

Ground emission can seriously affect continuum observations,  
especially in our halo fields where the emission has a brightness of only a few~mK. 
Our tests have shown that the highest ground emission 
occur in the Zenith cap at elevations above $60^\circ$
where large fluctuations are observed in the data. 
Even though not yet fully understood, the most likely reason is the loss of 
ground shielding by the upper rim of the dish at large elevations. The receiver is located at the 
prime focus and is shielded by the dish up to this elevation, receiving 
only atmospheric contributions from beyond the upper rim. 
Above this limit the ground becomes visible, contributing a ground component that rapidly varies as the telescope scans.
To avoid this contamination all PGMS observations were limited to the elevation 
range EL~=~$[30^\circ, 60^\circ]$,  between the lower limit of the telescope's motion and this region of ground sensitivity.

Even though these precautions have significantly reduced the effect, 
some contamination is still present in the halo fields, requiring us to develop 
a procedure to estimate and clean the ground contribution. 

\subsection{Estimate and cleaning procedure}

The procedure is based on making a map of the ground emission in 
the AZ--EL reference frame. 
Any AZ--EL bin gathers the contributions of data taken at different 
Galactic coordinates and the weak sky emission is efficiently averaged out. 
Since the PGMS meridian goes through low emission regions, its high latitude data are ideal 
for such an aim.
The smooth behaviour of the ground emission enables the use of large
bins, which further helps average out the sky component.   We
therefore use a bin size of $\Delta$EL = 1deg in EL and average over 8
degrees in Azimuth. The binning is performed in the instrument
reference frame before the correction for parallactic angle.
\begin{figure}
\centering
  \includegraphics[angle=00, width=1.0\hsize]{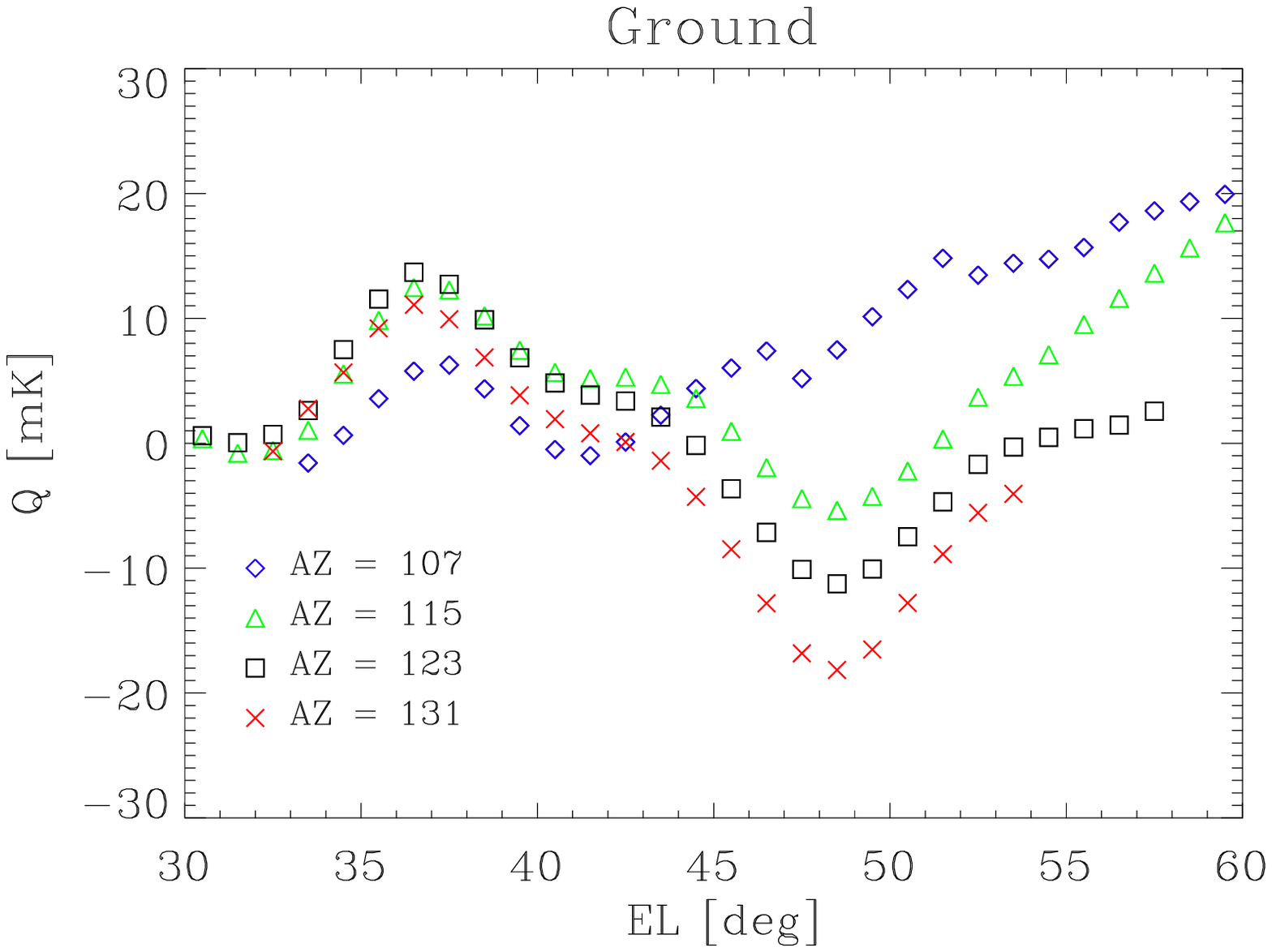}
  \includegraphics[angle=00, width=1.0\hsize]{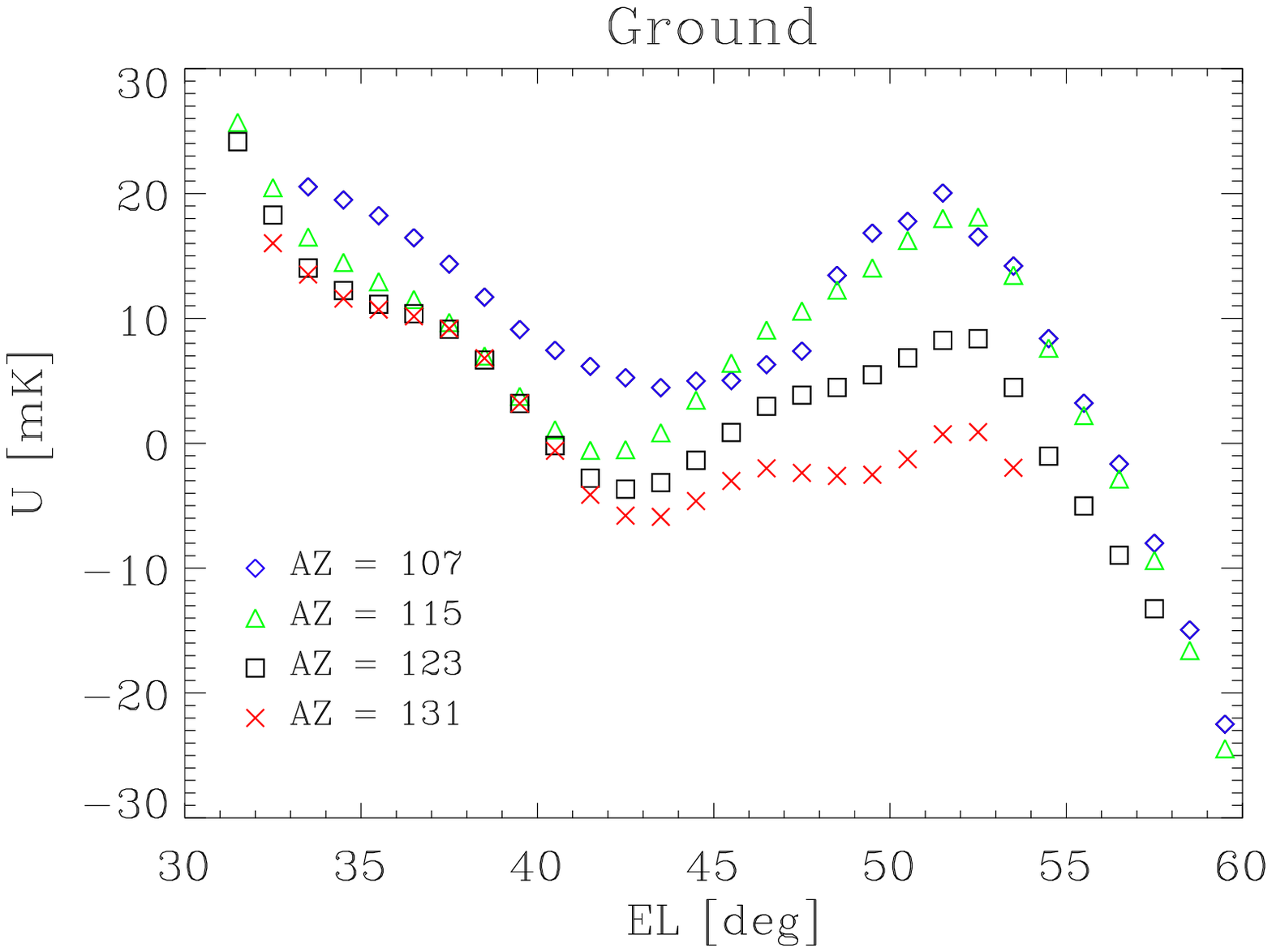}
\caption{Examples of azimuth cuts of the ground emission maps of both $Q$ (top) and $U$ (bottom)
                from the 2300~MHz channel of the September 2007 observations. 
\label{grnd_em_map:Fig}
}
\end{figure}

For a given 8 degree AZ bin we smooth the map along the EL
direction. This reduces the residual local deviations that are mainly
due to strong point sources. We use a quadratic running fit: for each
EL bin, we fit the 7 bins centred at it with a 2-degree polynomial,
and the bin value is then replaced with the fit result at the bin position.

In the Azimuthal direction the data are sufficiently smooth that no
fit is necessary in that dimension.   We therefore shift our 8 degree
azimuth averages by 1 degree increments, performing the elevation fit
for each 1 degree bin.  This results in a map of the ground emission
in the AZ-EL frame with a bin size of 1 degree in both Azimuth and
Elevation.
 
The ground emission contamination mainly comes from the far lobes,
which are frequency dependent. We generate ground emission maps 
for each frequency channel.
A set of these maps is generated at each observing session.   They are
checked for constancy of ground emission before a grand average
set is formed from all observations.  

Fig.~\ref{grnd_em_map:Fig} shows example AZ cuts in the 2300~MHz
spectral channel. Over a range of about $30\degr$ in both AZ and EL the
ground emission varies by less than 50~mK, 
smaller by an order of magnitude than ground emission variations reported 
for other polarization surveys (e.g. \citealt{wolleben06}). We attribute
this low response to ground emission to the high edge tapering of the
S-band feed and its consequent small sidelobes.

Finally, the sky emission measurements, observed on a 3-arcminute
grid, were cleaned of ground emission using the model just described. 
For each sky measurement, the ground emission at its actual (AZ, EL)
was obtained 
by linear interpolation through a standard Cloud-In-Cell technique~\citep{hockney81}.

\subsection{Cleaning procedure tests}
The low emission fields in the halo have been observed twice giving  
two independent maps taken in different runs and at different AZ and EL.
That way, they are contaminated by different ground emission enabling us 
to test the cleaning procedure and estimate the residual contamination.
The error map can be estimated as half the difference of the two maps,
in which the sky is cancelled, leaving the noise and any residual
(unmodelled) ground emission.

\begin{figure*}
\centering
  \includegraphics[angle=00, width=0.49\hsize]{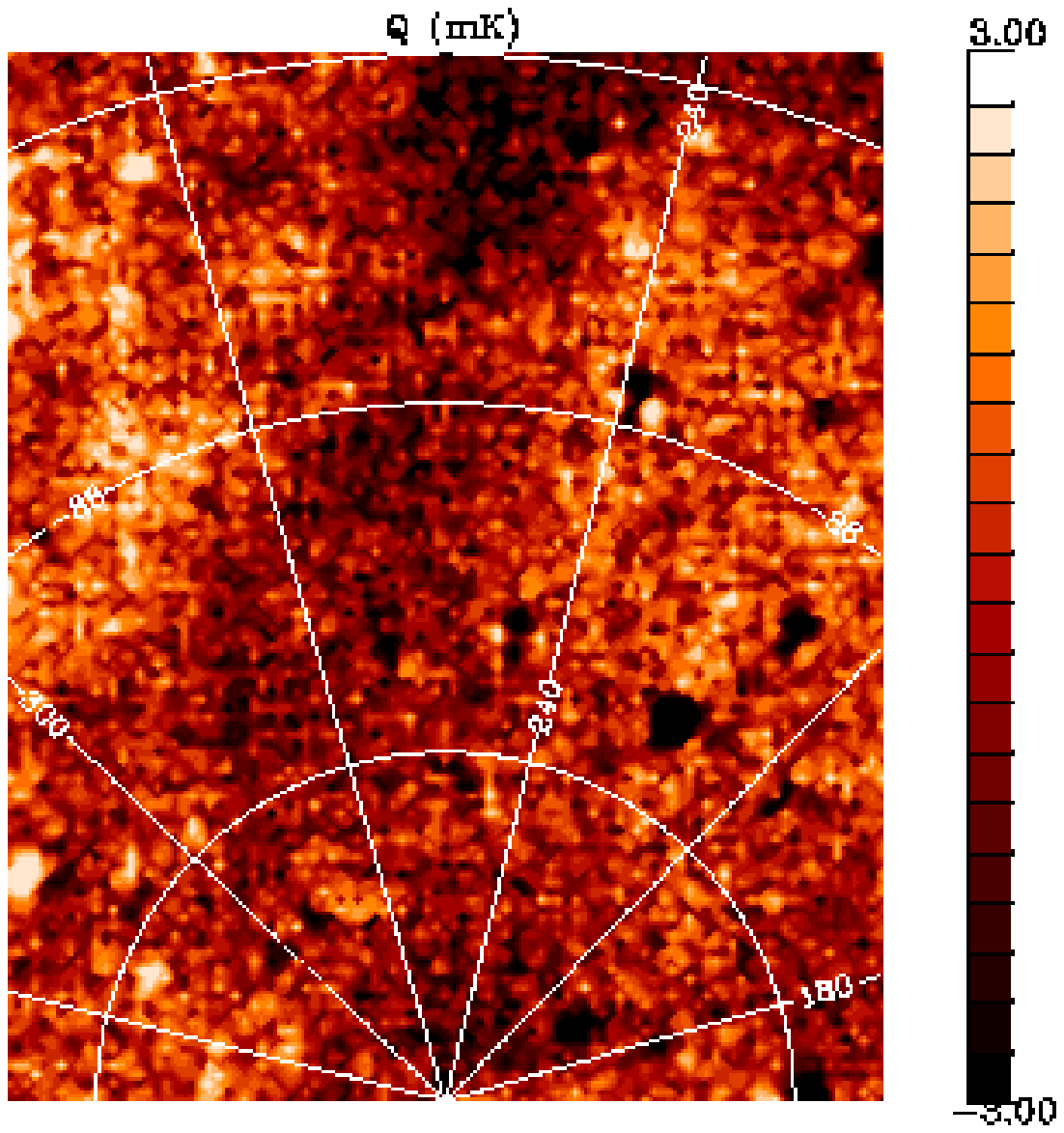}
  \includegraphics[angle=00, width=0.49\hsize]{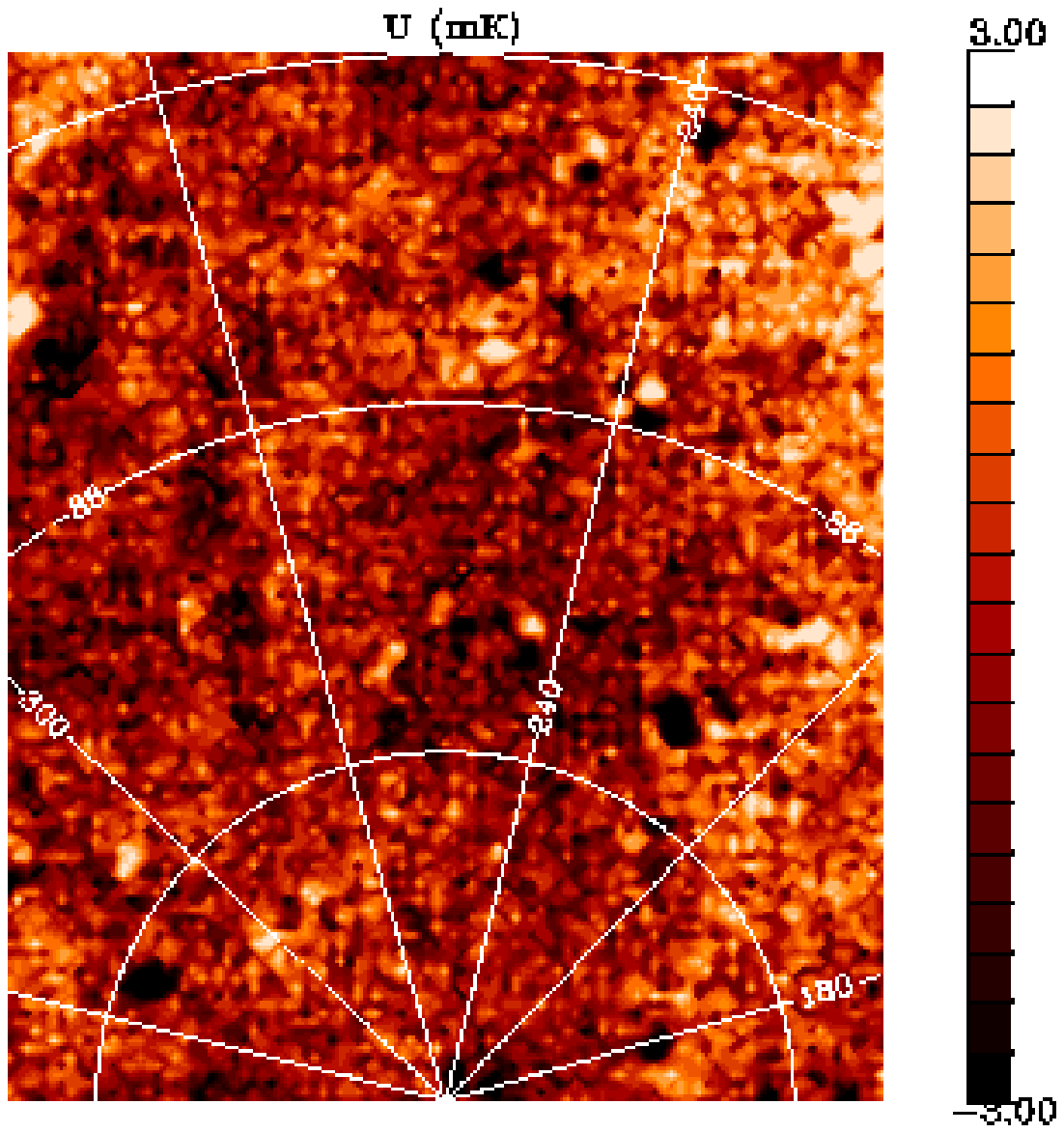}
  \includegraphics[angle=00, width=0.49\hsize]{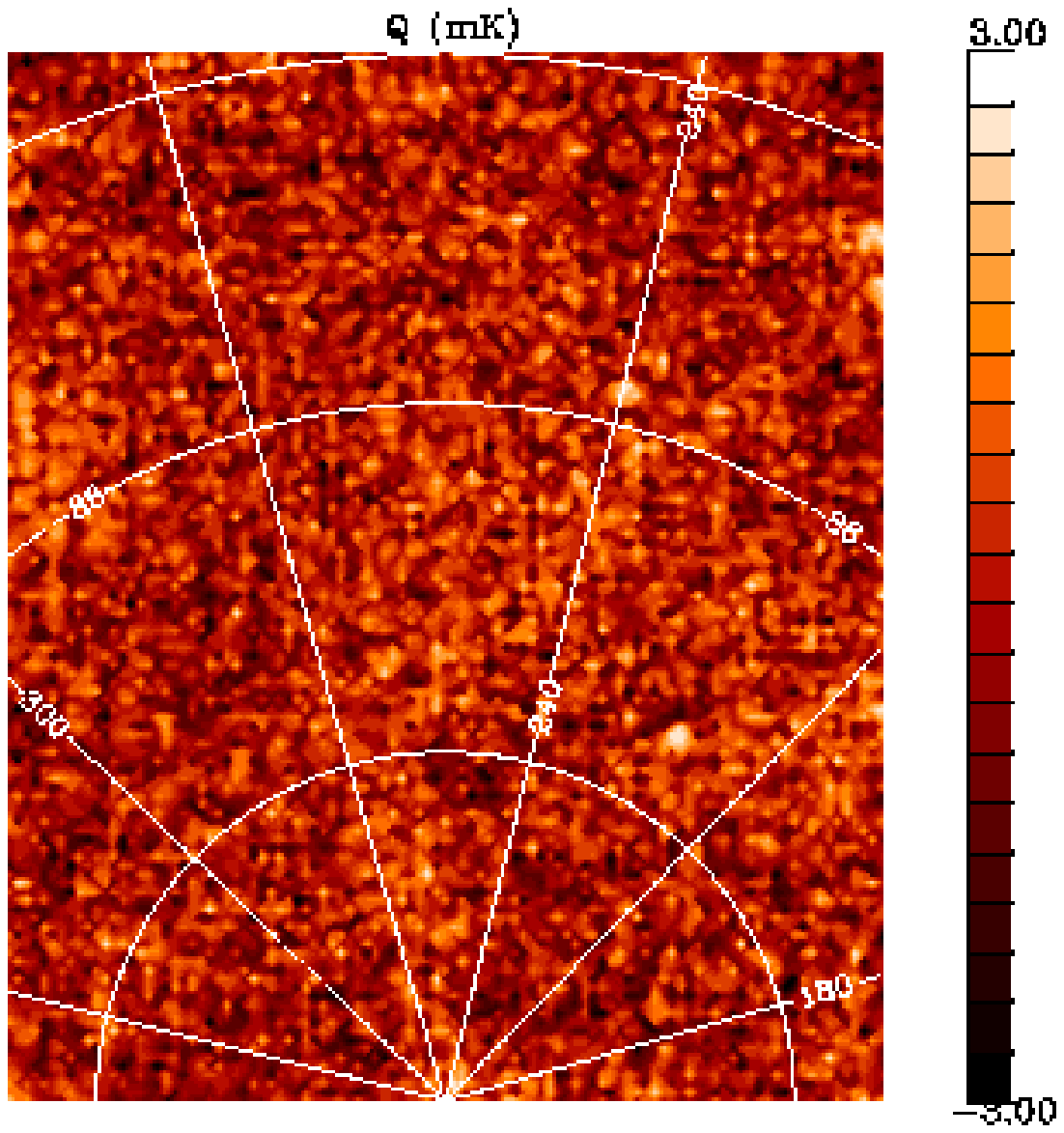}
  \includegraphics[angle=00, width=0.49\hsize]{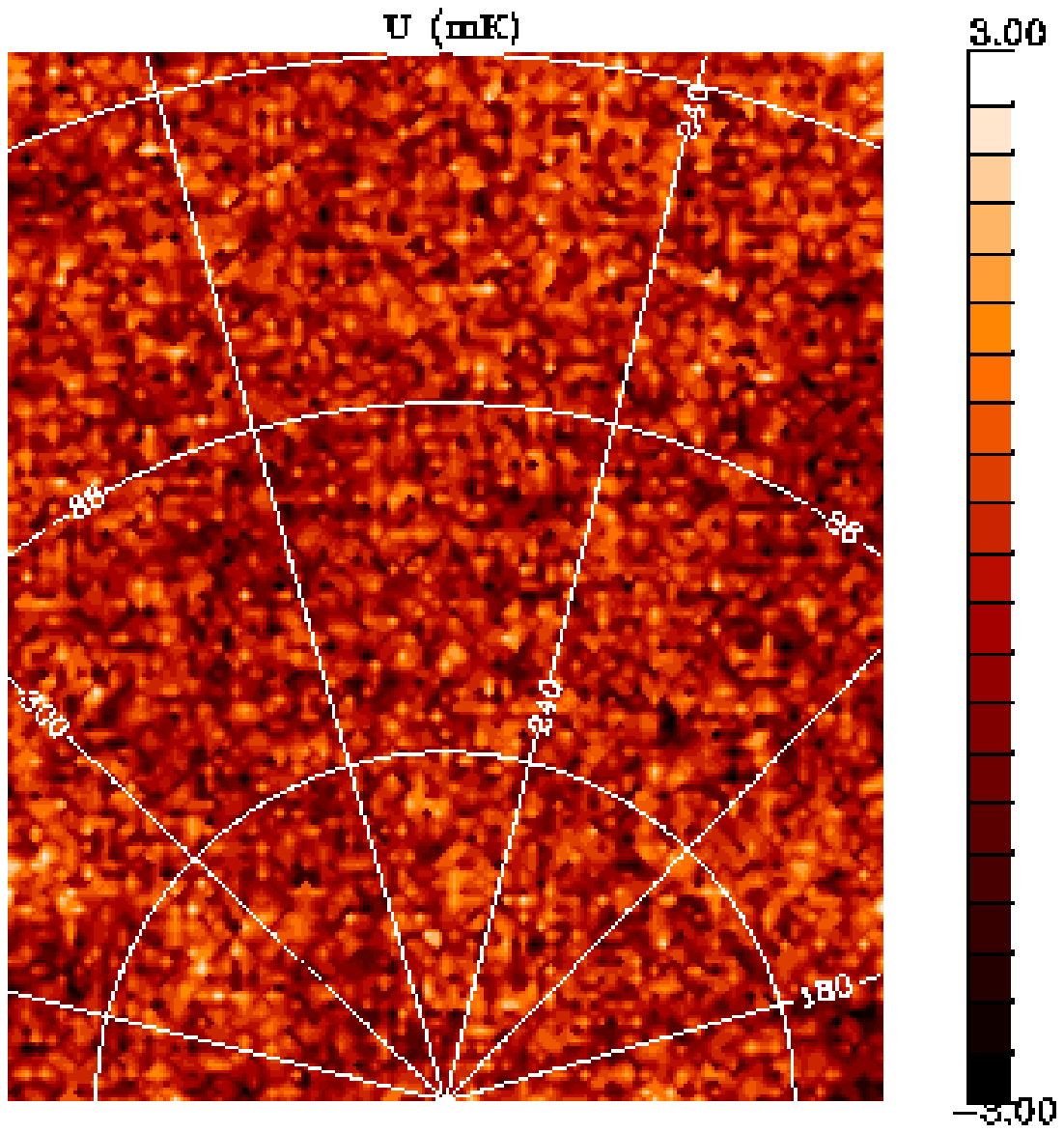}
\caption{ $Q$ (left) and $U$ images (right) of the observed (top) and difference maps (bottom)
                 of the field PGMS-87 ($b$=[-84$^\circ$,~-90$^\circ$]). No smoothing is applied 
                 for a resolution of FWHM=8.9'. 
                 Position coordinates are Galactic latitude and longitude; the brightness unit is mK, 
                 the intensity colour scales are linear.
\label{errorMaps:Fig}
}
\end{figure*}

An example is given in Fig.~\ref{errorMaps:Fig} in which one of the
fields with lowest emission is shown: PGMS-87. 
The images show Stokes $Q$ (left) and $U$ (right) of both the observed (top) 
and difference maps (bottom).  

The difference maps are clearly dominated by white noise, indicating
that most of the ground emission has been removed at the level needed
to measure the sky signal, and that the residual ground emission does
not contribute significantly to the error budget.  That residual can
be seen as faint {\it shadows} of large scale structure in the difference maps.

\begin{figure}
\centering
\includegraphics[angle=0, width=1.0\hsize]{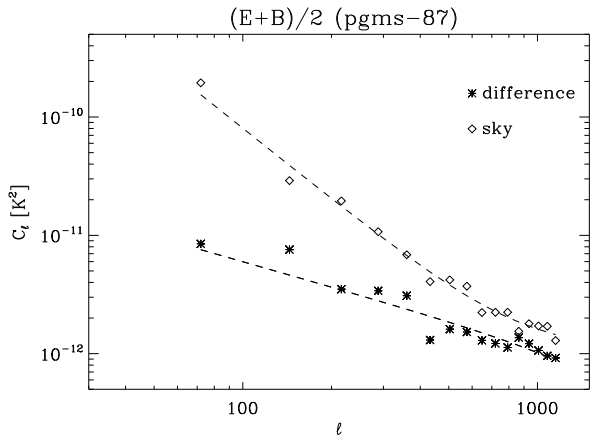}
\caption{  Power spectrum of the mean ($E$+$B$)/2 of $E$-- and $B$--Modes 
                  measured for the PGMS-87 polarized emission:
                  sky emission (diamonds), difference map (stars) and
                  best fit curves (dashed).
 \label{groundSpec:Fig}
}
\end{figure}

These visual impressions of the ground emission removal we know quantify
by measuring the angular power spectra\footnote{See Section~\ref{aps:Sect} 
for the description of the power spectrum computation.}
of both the sky and the difference map.  
Fig.~\ref{groundSpec:Fig} reports the mean of the $E$-- and $B$--mode power spectra 
-- ($E$+$B$)/2 --
which is the most complete description of the polarized emission. 

The spectrum of the sky signal is dominated by the diffuse emission at low multipole $\ell$, where it
follows a power law $C_\ell \propto \ell^\beta$ with a steep slope ($\beta < -2.0$).
A flattening occurs at the high--$\ell$ end due to both noise and a 
point source contribution. 
A white noise spectrum would be flat ($C_\ell$ = constant).

The difference spectrum also follows a power law, but is much flatter than
the sky signal. 
The best fit slope is $\beta_{\rm noise} = -0.70\pm0.06$, which, although not pure white noise,
is close to the ideal  $\beta_{\rm noise} = 0$.
Furthermore, the difference 
between sky signal and noise increases at large angular scale,  giving
a rapidly increasing S/N. 

The ground emission has a smooth behaviour that makes the largest
scales the most susceptible to contamination.
On a scale of 2$^\circ$ the rms fluctuation of the difference map is $N_{2^\circ} = 60$~$\mu$K, 
a factor 2.5 larger than that expected from white noise (24~$\mu$K),
but much smaller than the sky signal (few~mK).

This field (PGMS-87) is the worst with regard to ground
emission residuals;  over all fields, angular spectral 
slopes $\beta_{\rm noise}$ fall in the range [-0.7, 0.0] and the rms noise on the 2-degree
scale $N_{2^\circ}$ lie in [24, 60]~$\mu$K. The mean  values over all
fields are $\bar{\beta}_{\rm noise} = -0.4$ and $\bar{N}_{2^\circ} =  
43$~$\mu$K. Once the pure white noise component is subtracted,  the effective  
contribution by only ground emission can be estimated in $\bar{N}_{{\rm grnd},2^ 
\circ} = 36$~$\mu$K. 
With such results the impact of the ground emission may be
considered marginal on the final mapping.

\section{PGMS maps}
\label{maps:Sect}

\begin{figure*}
\centering
  \includegraphics[angle=00, width=0.33\hsize]{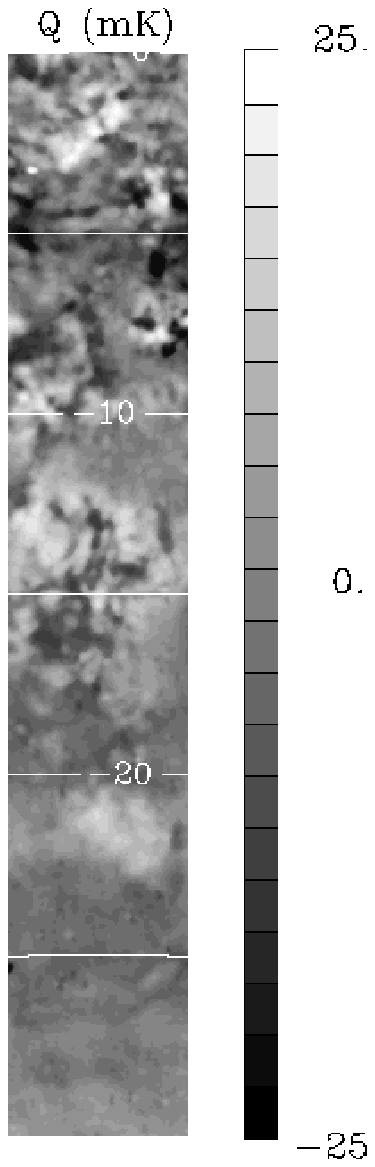}
  \includegraphics[angle=00, width=0.33\hsize]{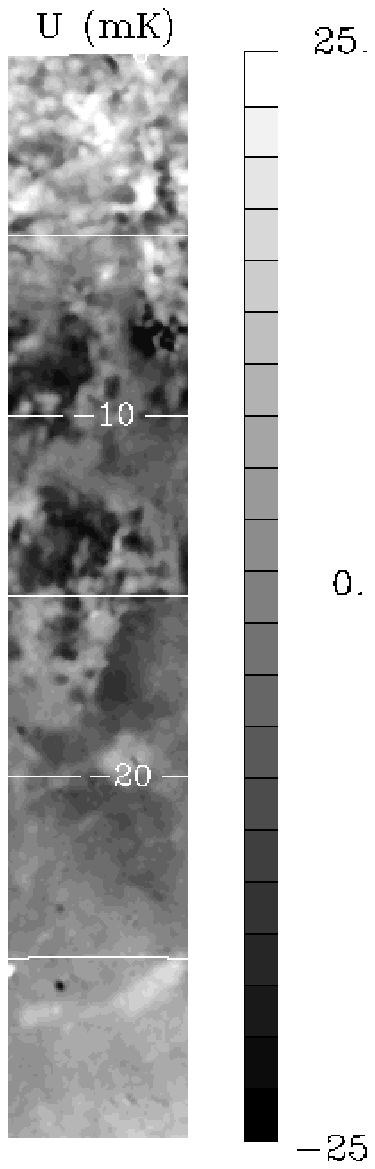}
  \includegraphics[angle=00, width=0.33\hsize]{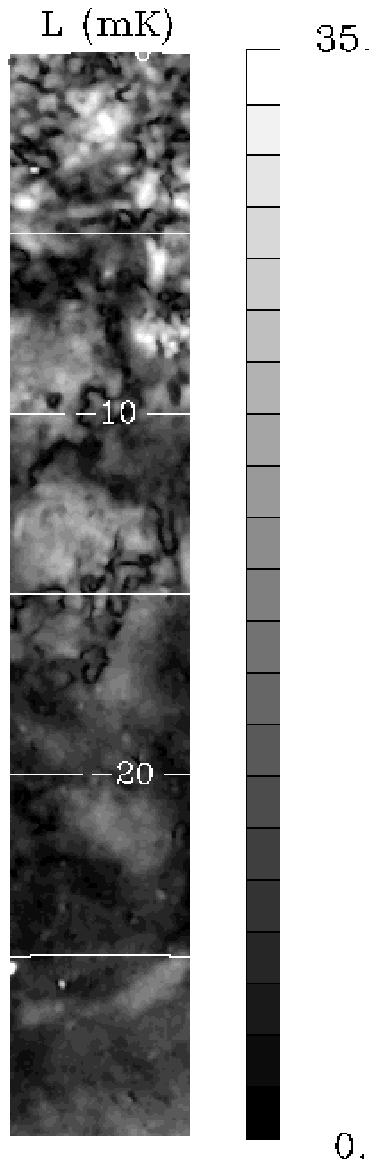}
\caption{ $Q$ (left), $U$ (mid), and polarized intensity $L=\sqrt{Q^2+U^2}$ images (right) 
                 of the six PGMS fields in the latitude range $b=[-30\degr, 0\degr]$ 
                 (PGMS-27 through PGMS-02).
                 Position coordinates are Galactic latitude; the brightness unit is mK.
                 \label{pgms-00-30:Fig}
}
\end{figure*}

\begin{figure*}
\centering
  \includegraphics[angle=00, width=0.33\hsize]{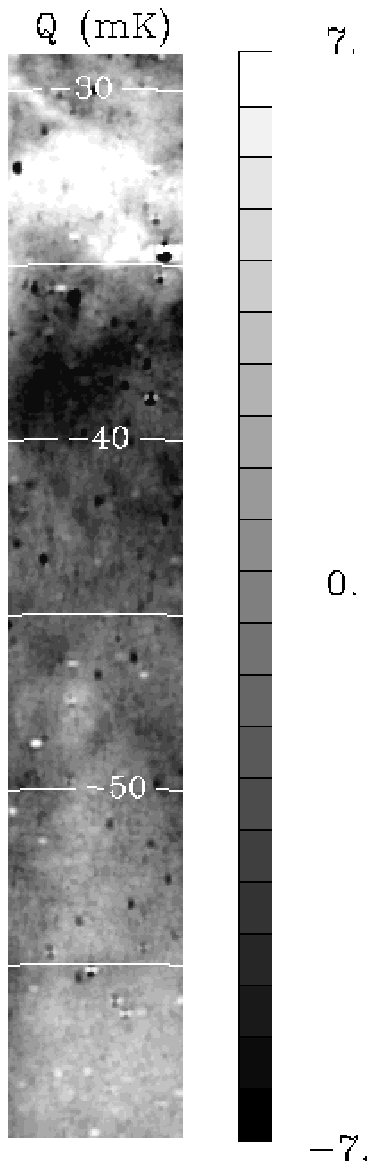}
  \includegraphics[angle=00, width=0.33\hsize]{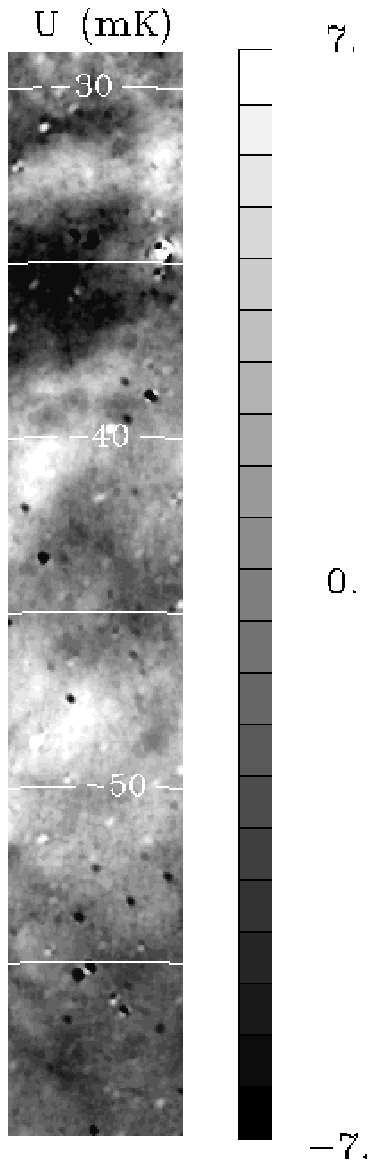}
  \includegraphics[angle=00, width=0.33\hsize]{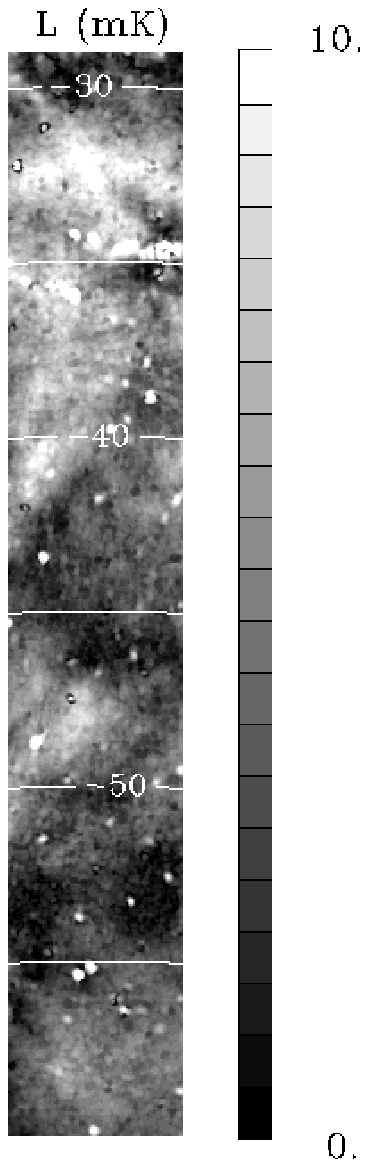}
\caption{ As for Fig.~\ref{pgms-00-30:Fig} but in the range $b=[-60\degr, -30\degr]$ 
                 (PGMS-57 through PGMS-34: of the latter only the $5\degr$ across the
                  meridian $l=254\degr$ are imaged).
                 Position coordinates are Galactic latitude; the brightness unit is mK.
\label{pgms-30-60:Fig}
}
\end{figure*}

\begin{figure*}
\centering
  \includegraphics[angle=00, width=0.33\hsize]{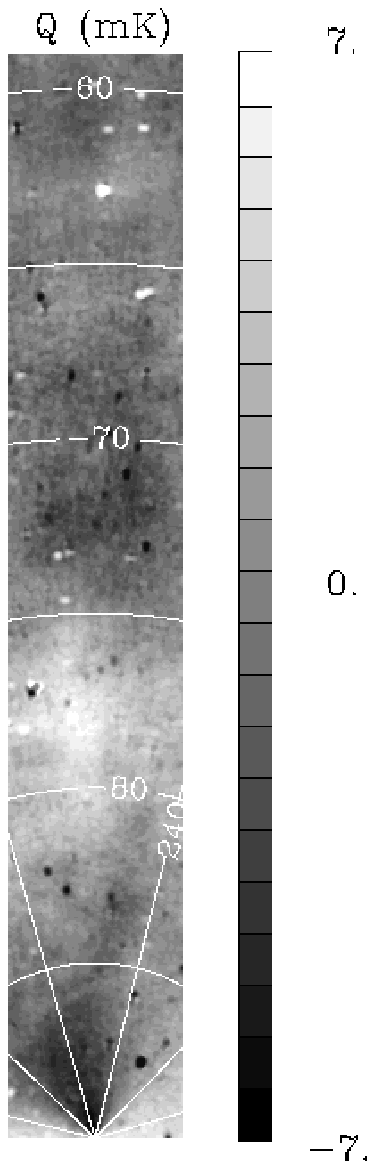}
  \includegraphics[angle=00, width=0.33\hsize]{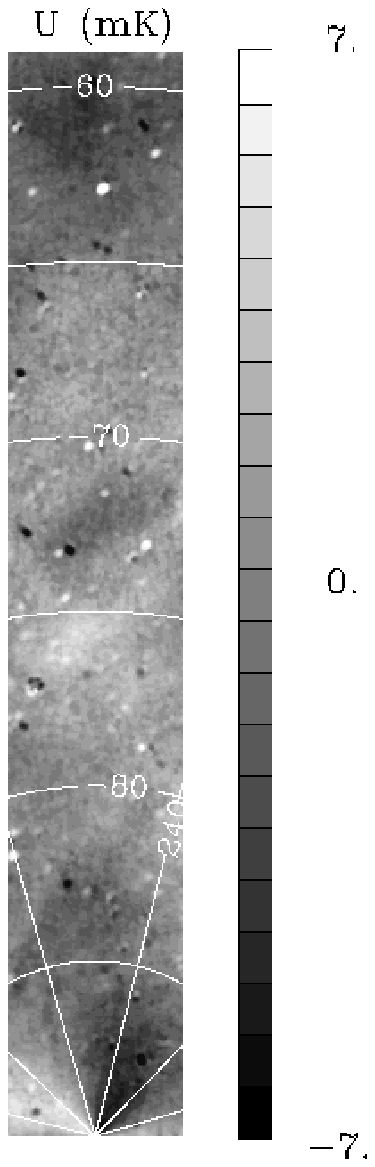}
  \includegraphics[angle=00, width=0.33\hsize]{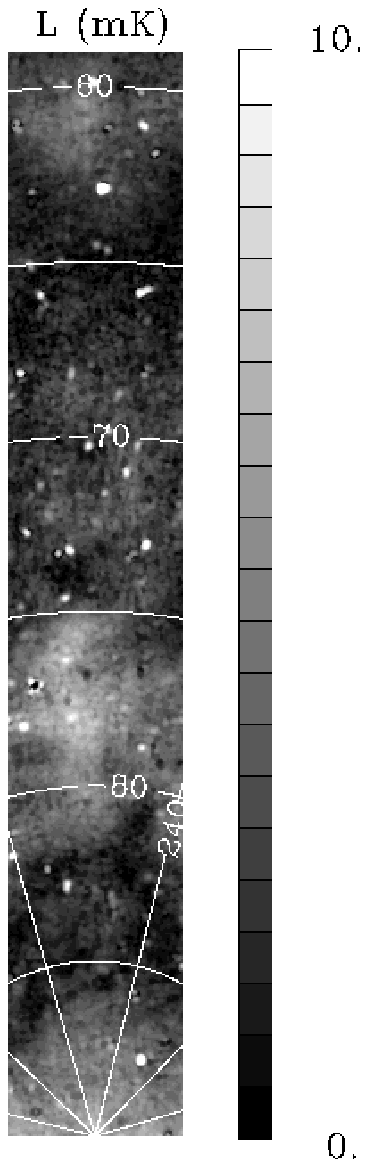}
\caption{ As for Fig.~\ref{pgms-00-30:Fig} but in the range $b=[-90\degr, -60\degr]$ 
                 (PGMS-87 through PGMS-62).
                 Position coordinates are Galactic latitude and longitude; the brightness unit is mK.
\label{pgms-60-90:Fig}
}
\vskip 1cm
\end{figure*}

\begin{figure*}
\centering
  \includegraphics[angle=00, width=0.55\hsize]{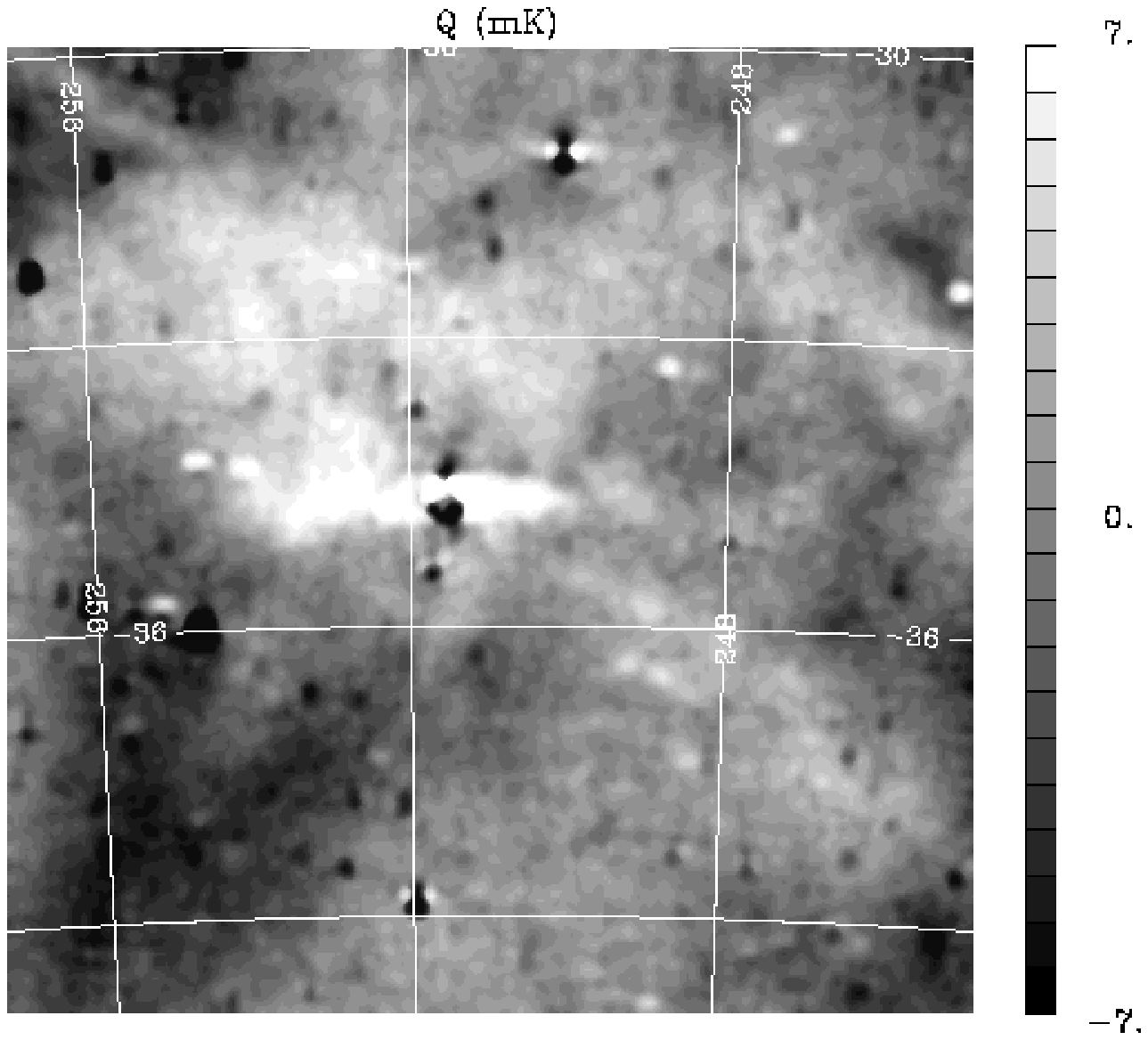}
  \includegraphics[angle=00, width=0.55\hsize]{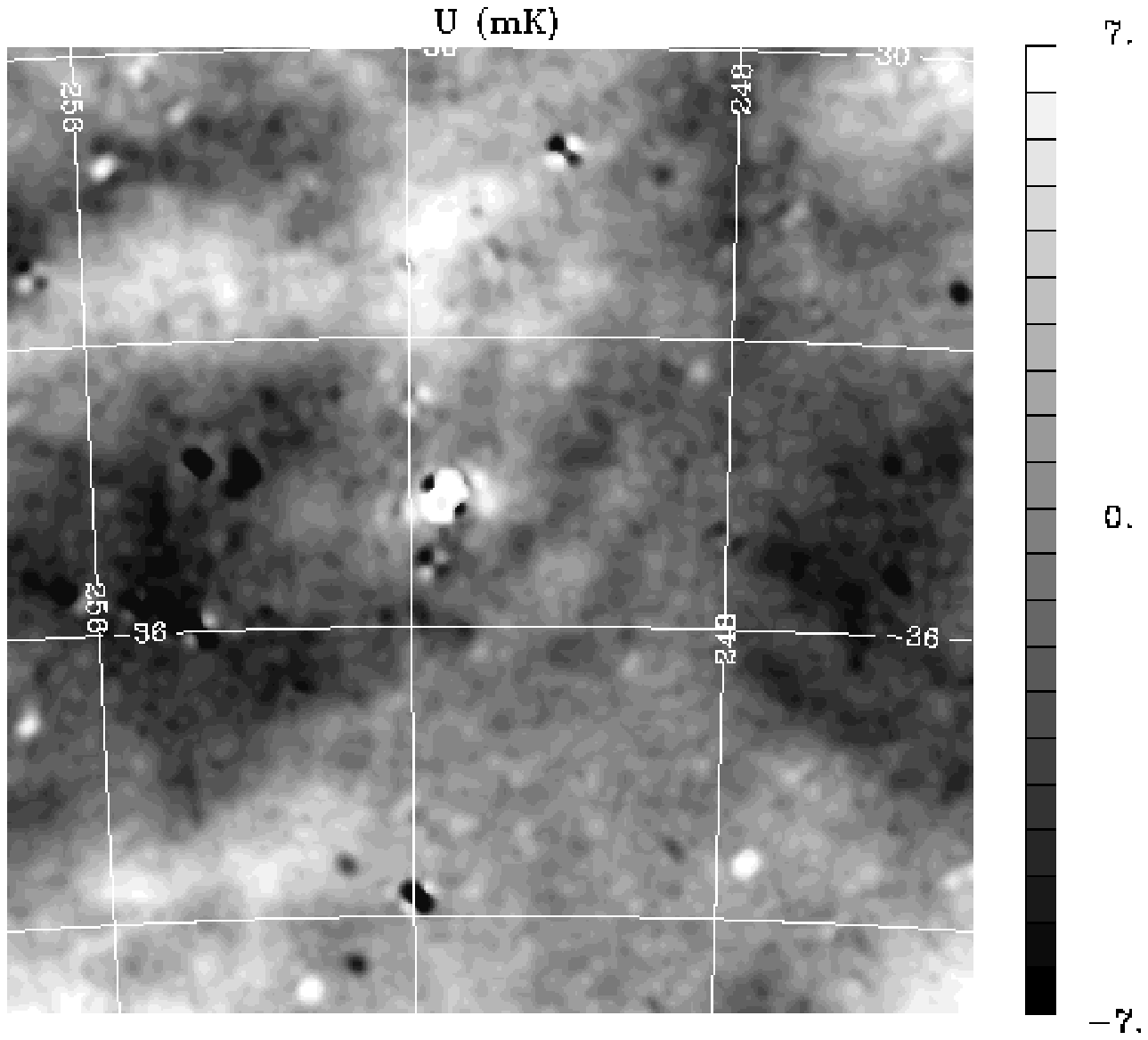}
  \includegraphics[angle=00, width=0.55\hsize]{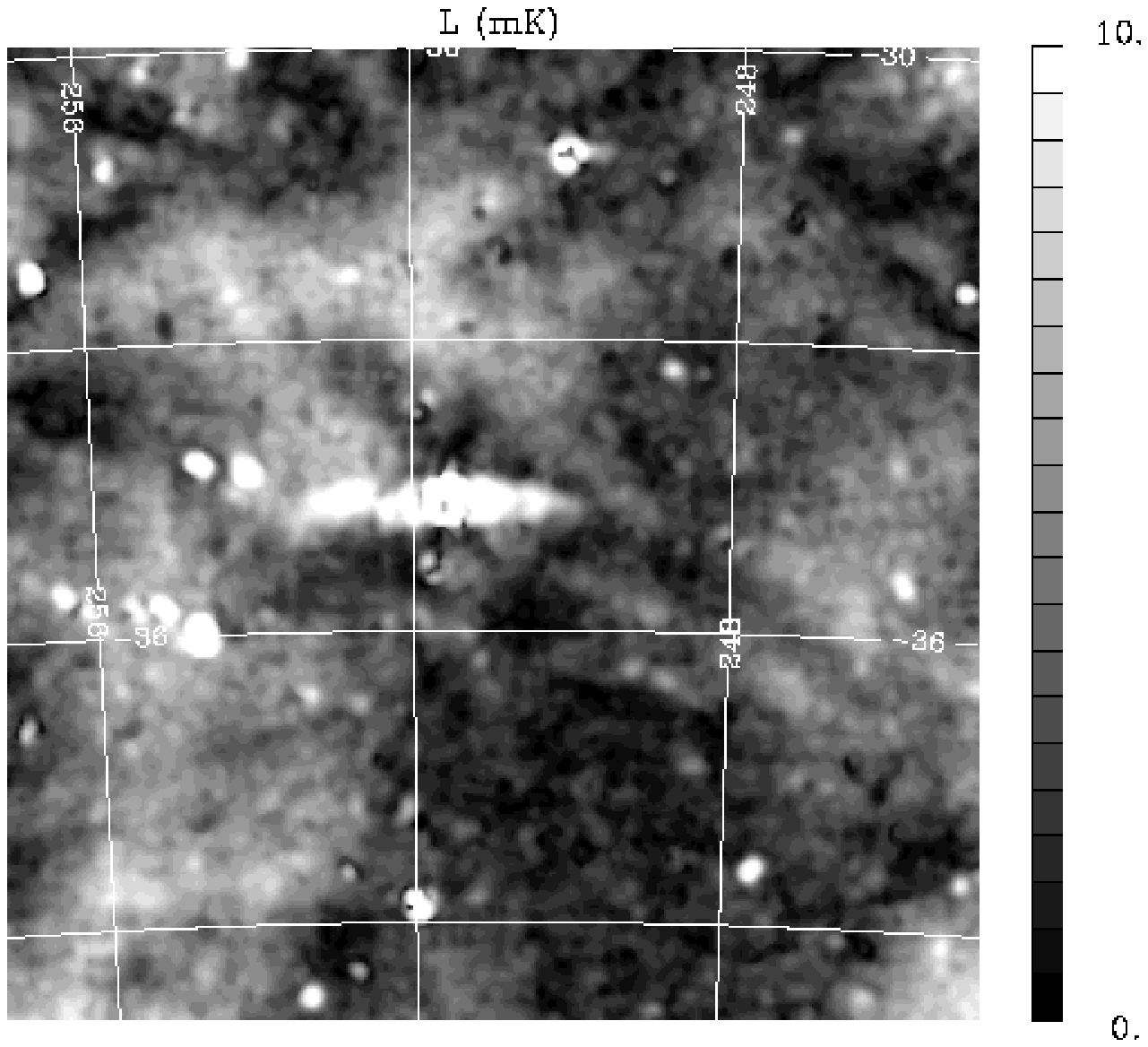}
\caption{ $Q$ (top), $U$ (mid), and $L$ images (bottom) 
                 of the whole $10\degr \times 10\degr$ fields (PGMS-34).
                 Position coordinates are Galactic latitude and longitude; the brightness unit is mK.
\label{pgms-34:Fig}
}
\end{figure*}

Maps of the Stokes parameters $Q$, $U$ and the polarized intensity $L$ 
of the PGMS meridian are shown in Fig.~\ref{pgms-00-30:Fig},~\ref{pgms-30-60:Fig},
and~\ref{pgms-60-90:Fig}, while Fig.~\ref{pgms-34:Fig} displays the whole $10\degr \times 10\degr$ 
field (PGMS-34). 
All images are smoothed with a Gaussian filter of FWHM~$=6\arcmin$
to give a better idea of the sensitivity on beam-size scale, for an effective 
resolution of FWHM~$=10.7\arcmin$. All data at latitude $|b| > 30^\circ$
are plotted with the same intensity range to show clearly the power and morphological differences. 
The disc fields ($|b| < 30^\circ$) require a more extended scale. 
The two strongest sources present in our data (Pic~A and NGC~612 in field PGMS-34
and PGMS-77, respectively) have been blanked before the map generation of their fields. Without blanking,
the high brightness range causes the map-making procedure to generate artefacts. 

The disc has the strongest emission, extending to latitude $|b|\sim 20\degr$ with little variation of emission power. 
At higher latitudes, the emission starts to decrease up to the halo where it settles
on levels one order of magnitude lower.

The clear visibility of the bright polarized disc emission and its
contrast with the fainter halo is a new result, 
not apparent from previous observations carried out at lower frequencies where the disc
 is strongly depolarized up to $|b| \sim 30\degr$. 
This allows us to locate the disc-halo boundary in polarization, which
has not been visible so far because of either strong depolarization 
(at 1.4~GHz) or insufficient sensitivity (at 23~GHz).
A more quantitative analysis is given in Section~\ref{aps:Sect}, but the visual inspection
of the maps clearly shows that it starts at $|b|\sim 20\degr$.

The emission of the halo has a smooth behaviour with the power mostly residing on 
large angular scales. The disc emission  is also smooth, at least at latitudes 
higher than $|b| = 6\degr$--$7\degr$. Closer to the Galactic plane 
the pattern has a more patchy appearance, suggestive of Faraday
depolarization effects being significant at 2.3~GHz.

This supports the view that Faraday depolarization effects are marginally
significant in the disc, and are relevant only in a narrow belt a few degrees
wide across the Galactic plane. 
\begin{table}
\centering
\caption{Polarization flux limits $S_p^{\rm lim}$ used to selected the
               polarized sources. The chosen limits are latitude dependent.
        }
\begin{tabular}{@{}cc@{}}
\hline
$b$-range  &  $S_p^{\rm lim}$~[mJy] \\
\hline
$[-10\degr,  0\degr]$   &  40 \\
$[-20\degr, -10\degr]$   &  30 \\
$[-30\degr, -20\degr]$   &  20 \\
$[-40\degr, -30\degr]$   &  15 \\
$[-90\degr, -40\degr]$   &  10 \\
\hline
\end{tabular}
\label{source_lim:Tab}
\end{table}

Several polarized point sources are visible, especially in the halo where the 
diffuse emission fluctuations are smaller. To enable a cleaner analysis 
of the diffuse component the sources are identified, fitted and subtracted 
from the maps.
Each source is located by a 2D-Gaussian fit of the stronger component, either $Q$ or $U$. 
Its position is then fixed in the fit of the second and weaker component 
to improve the fit robustness. 
A polarization flux limited selection is applied with threshold set to ensure $S/N$ 
ratios of at least 5. The amplitude of fluctuations in the maps is dominated 
by sky emission (rather than by the instrument sensitivity), which varies along 
the PGMS meridian.  The threshold we use is therefore a function of Galactic latitude, 
running from 10~mJy at high latitudes up to 40~mJy near the Galactic plane 
(Table~\ref{source_lim:Tab}).

In this work, the point source identification is carried out only for 
cleaning purposes. The catalogue and a detailed analysis
of their properties are subject of a forthcoming paper (Bernardi et al. 2010, in preparation).

\section{Angular power spectra}
\label{aps:Sect}

The angular power spectra (APS) of $E$-- and $B$--Mode of the 
polarized emission have been computed for each field. 
They account for the 2-spin tensor nature of the polarization
and give a full description of the polarized signal and its behaviour
across the range of angular scales. In addition, the $E$-- and $B$--Modes 
are the quantities predicted by the cosmological models enabling 
a direct comparison with the CMB. 

To cope with both the non-square geometry and the blanked pixels at the locations of
the two brightest sources, we use a method based on the two-point 
correlation functions of the Stokes parameters $Q$ and $U$  
described by \citet{sbarra03}. 
The correlation functions are estimated on the $Q$ and $U$ maps 
of the regions as   
\begin{equation}     
 \tilde{C}^X(\theta) = X_i X_j \hspace{1cm} X =    
 Q,U
\end{equation}     
where $X_i$ is the emission in pixel $i$ of map $X$, and $i$ and $j$    
identify pixel pairs at distance $\theta$.
Data are binned with pixel-size resolution.
Power spectra $C^{E,B}_\ell$ are obtained by integration 
\begin{equation}      
  C^E_\ell = W_\ell^P 
              \int^\pi_0 [\tilde{C}^Q(\theta)F_{1,\ell 2}(\theta) +     
                          \tilde{C}^U(\theta)F_{2,\ell 2}(\theta)]   
                          \sin(\theta)d \theta
\end{equation}      
\begin{equation}      
 C^B_\ell  = W_\ell^P 
              \int^\pi_0 [\tilde{C}^U(\theta)F_{1,\ell 2}(\theta) +     
                           \tilde{C}^Q(\theta)F_{2,\ell 2}(\theta)]   
                           \sin(\theta)d \theta 
\end{equation}
where $F_{1,\ell m}$ and $F_{2,\ell m}$ are functions of 
Legendre polynomials (see \citealt{zaldarriaga98} for their definition), 
and $W_\ell^P$ is the pixel window function
accounting for pixel smearing effects.

Since the emission power is best described by the quantity 
$\ell(\ell+1)C_\ell/(2\pi)$,  hereafter we will denote 
an angular spectrum following a power law behaviour $C_\ell \propto \ell^\beta$ as
\begin{itemize}
  \item{} {\it flat}, if $\beta = -2.0$: power equally distributed across the angular scales;
  \item{} {\it steep}, if $\beta < -2.0$: large scales dominate the power budget;
  \item{} {\it inverted}, if $\beta > -2.0$: small scales dominates.
\end{itemize}

We tested the procedure using simulated maps generated from a known 
input power spectrum by the procedure {\it synfast}  of the software package
HEALPix~\citep{go05}.
The input spectra are power laws with different slopes; for each slope
we generated 100 simulated maps and compute their APS. 
The mean of the 100 APS reproduced the input spectrum
and its slope correctly, with the exception of an excess 
at the largest scales, mainly at the first two multipole bands. 
$E$-- and $B$--Mode are related to the polarization angle pattern and 
this excess is likely due to the discontinuity of the pattern 
 abruptly interrupted at the area borders.
To account for this we corrected our spectra for the fractional 
excess estimated from the simulation as the ratio between 
the mean of the computed and input spectra. 

For a cleaner measure of the diffuse component, the point sources 
are subtracted from the polarization maps. 

The $E$-- and $B$--mode spectra $C^E_\ell$ and $C^B_\ell$ have been computed for the 
17~fields along with their mean $C_\ell^{(E+B)/2} = (C_\ell^E+C_\ell^B)/2$. 
Artificial fluctuations are generated on $E$ and $B$ spectra because of
the limited sky coverage of the individual areas, but 
their mean suffers less from that effect and is a more accurate estimator 
if the power is distributed equally between the two modes, as is the
case for Galactic emission.
In addition, $C^{(E+B)/2}_\ell$ gives a full description of the polarized emission 
which the two individual spectra cannot give separately. 
Therefore we mostly use the mean spectrum ($E$+$B$)/2 to investigate 
emission behaviour and properties in the following analysis. 

\begin{figure*}
\centering
  \includegraphics[angle=0, width=1.0\hsize]{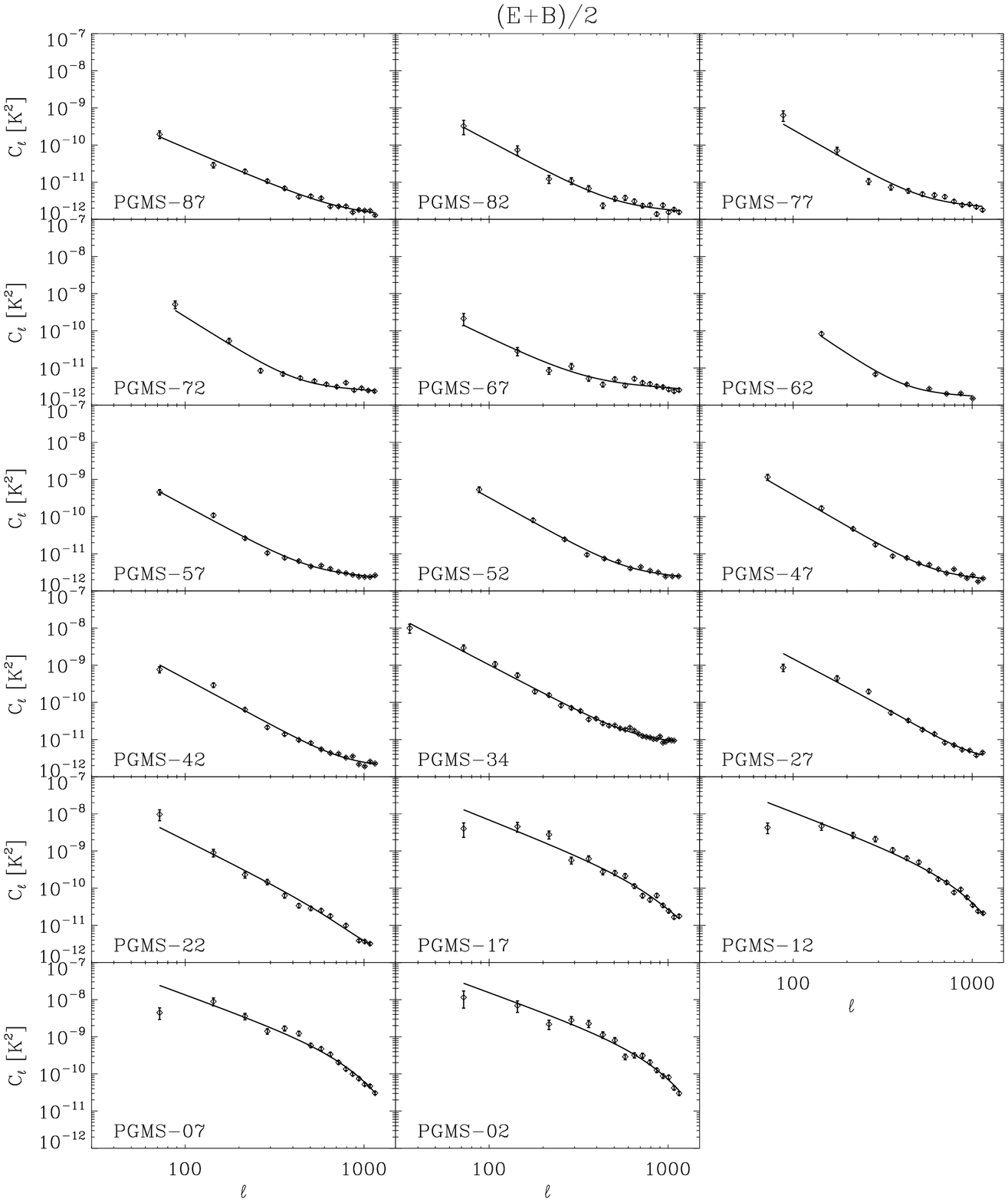}
\caption{ Angular power spectra $C_\ell^{(E+B)/2}$ of the 17~PGMS fields. Both the measured spectra (diamonds)
                and the best fit curve (solid) are plotted.
                \label{all_meas_specs:Fig}
}
\end{figure*}

\begin{figure*}
\centering
  \includegraphics[angle=0, width=0.45\hsize]{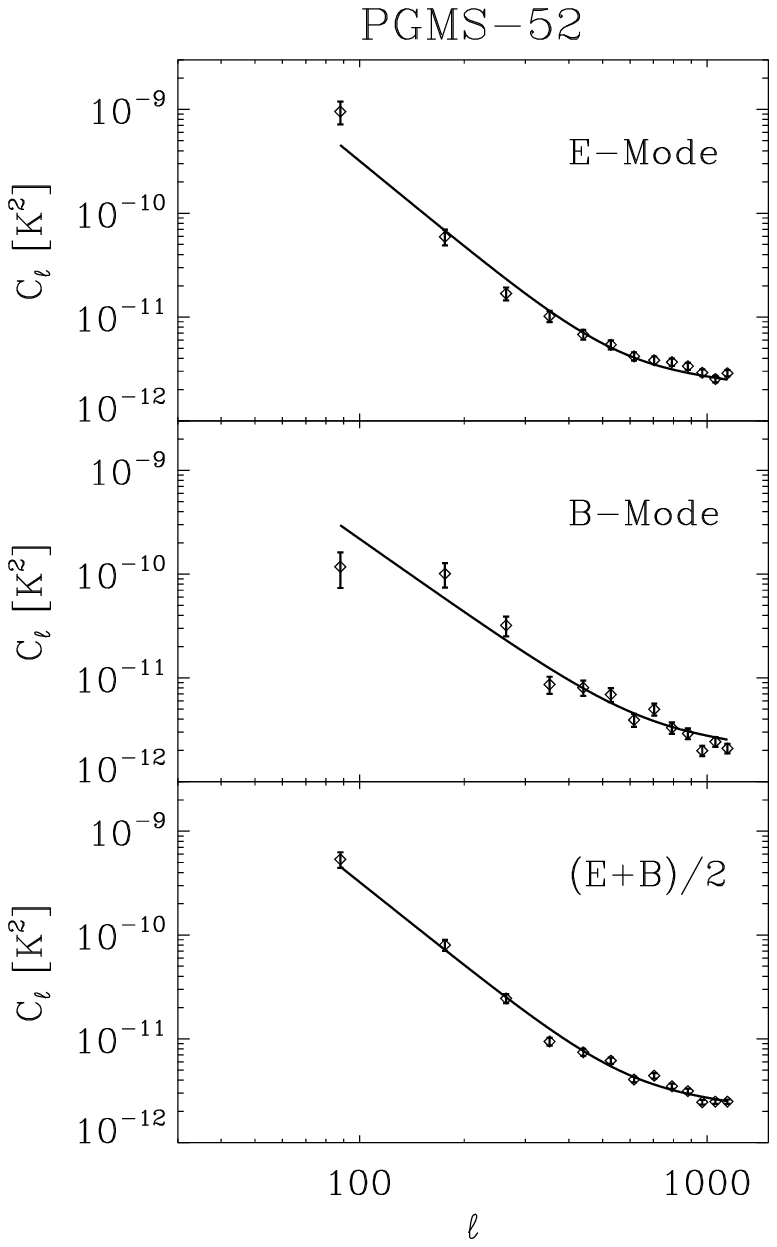}
  \includegraphics[angle=0, width=0.45\hsize]{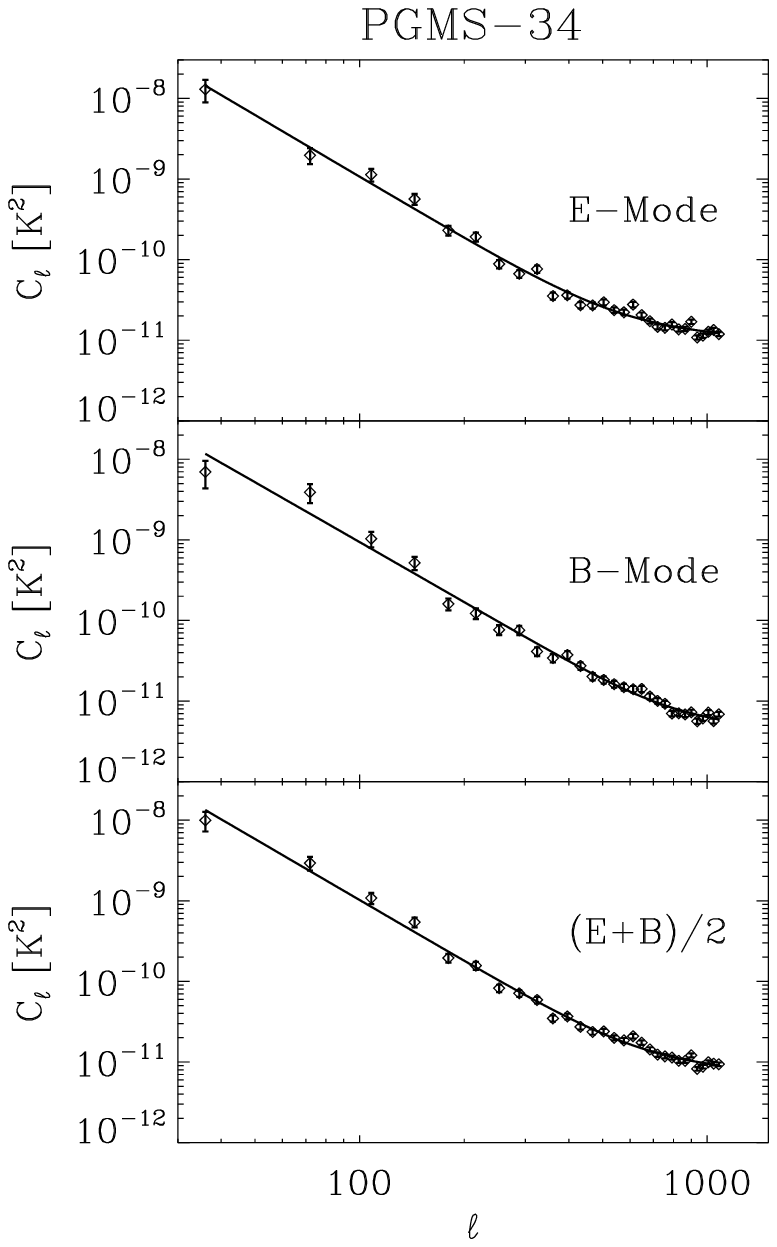}
\caption{Angular power spectrum of $E$--Mode (top), $B$--Mode (mid), and their mean ($E$+$B$)/2 of 
                the two fields PGMS-52 (left) and PGMS-34 (right). Both the measured spectra (diamonds)
                and the best fit curve (solid) are plotted.
                \label{specs_E_B_EB:Fig}
}
\end{figure*}

Fig.~\ref{all_meas_specs:Fig} shows $C^{(E+B)/2}_\ell$ for all the
fields.  As an example of all three spectra,
Fig.~\ref{specs_E_B_EB:Fig} shows those of the
two fields PGMS-52, which is from the low emission halo, and PGMS-34,
our biggest field and the area observed by the BOOMERanG experiment.
All spectra are shown without correction for the window
functions.
\begin{table*}
\centering
\caption{Best fit amplitude $C^X_{200}$ (referenced to $\ell=200$)
               and angular spectral slope $\beta^X$ 
                of the PGMS fields ($X=E, B, (E+B)/2$ denoting $E$--, $B$--Mode, and ($E$+$B$)/2,
                respectively). 
                \label{bestFit:Tab} 
}
\begin{tabular}{@{}lcccccc@{}}
\hline
 Field  &  $C^E_{200}$~[$\mu$K$^2$]   &  $\beta^E$  &  
           $C^B_{200}$~[$\mu$K$^2$]   &  $\beta^B$  &  
           $C^{(E+B)/2}_{200}$~[$\mu$K$^2$]  &  $\beta^{(E+B)/2}$  \\ 
\hline
 PGMS-02 & $     4500 \pm      280 $ & $    -1.69 \pm     0.04 $ & $     4860 \pm      880 $ & $    -1.90 \pm     0.13 $ & $     4610 \pm      540 $ & $    -1.76 \pm     0.09 $  \\
 PGMS-07 & $     3850 \pm      380 $ & $    -1.77 \pm     0.07 $ & $     4120 \pm      340 $ & $    -1.75 \pm     0.06 $ & $     4030 \pm      340 $ & $    -1.76 \pm     0.06 $  \\
 PGMS-12 & $     2800 \pm      200 $ & $    -1.86 \pm     0.05 $ & $     3160 \pm      410 $ & $    -1.82 \pm     0.09 $ & $     3070 \pm      240 $ & $    -1.85 \pm     0.06 $  \\
 PGMS-17 & $     2090 \pm      260 $ & $    -2.00 \pm     0.10 $ & $     1270 \pm      140 $ & $    -1.67 \pm     0.08 $ & $     1840 \pm      190 $ & $    -1.91 \pm     0.08 $  \\
 PGMS-22 & $      550 \pm      100 $ & $    -2.54 \pm     0.17 $ & $      246 \pm       39 $ & $    -2.60 \pm     0.16 $ & $      376 \pm       31 $ & $    -2.39 \pm     0.08 $  \\
 PGMS-27 & $      287 \pm       58 $ & $    -2.41 \pm     0.24 $ & $      131 \pm       19 $ & $    -2.17 \pm     0.19 $ & $      263 \pm       20 $ & $    -2.51 \pm     0.09 $  \\
 PGMS-34 & $    183.5 \pm      8.5 $ & $    -2.54 \pm     0.07 $ & $    171.8 \pm      9.0 $ & $    -2.46 \pm     0.07 $ & $    180.0 \pm      7.1 $ & $    -2.51 \pm     0.06 $  \\
 PGMS-42 & $       79 \pm       11 $ & $    -2.76 \pm     0.19 $ & $     61.9 \pm      6.3 $ & $    -2.24 \pm     0.14 $ & $     74.1 \pm      4.0 $ & $    -2.57 \pm     0.08 $  \\
 PGMS-47 & $     59.8 \pm      7.7 $ & $    -2.91 \pm     0.20 $ & $     49.7 \pm      4.3 $ & $    -2.56 \pm     0.13 $ & $     56.3 \pm      3.1 $ & $    -2.78 \pm     0.08 $  \\
 PGMS-52 & $     47.9 \pm      4.4 $ & $    -2.74 \pm     0.13 $ & $     42.7 \pm      5.9 $ & $    -2.36 \pm     0.17 $ & $     51.3 \pm      3.2 $ & $    -2.66 \pm     0.08 $  \\
 PGMS-57 & $     46.7 \pm      3.4 $ & $    -2.62 \pm     0.10 $ & $     16.6 \pm      2.2 $ & $    -2.54 \pm     0.21 $ & $     32.4 \pm      1.7 $ & $    -2.61 \pm     0.07 $  \\
 PGMS-62 & $     28.9 \pm      3.1 $ & $    -3.27 \pm     0.22 $ & $     19.1 \pm      3.4 $ & $    -3.35 \pm     0.40 $ & $     24.1 \pm      1.9 $ & $    -3.27 \pm     0.17 $  \\
 PGMS-67 & $     15.6 \pm      2.0 $ & $    -2.17 \pm     0.19 $ & $      5.4 \pm      1.1 $ & $    -2.23 \pm     0.34 $ & $     12.6 \pm      1.6 $ & $    -2.35 \pm     0.20 $  \\
 PGMS-72 & $     40.1 \pm      4.1 $ & $    -3.08 \pm     0.16 $ & $     14.9 \pm      2.8 $ & $    -2.81 \pm     0.33 $ & $     29.7 \pm      2.7 $ & $    -3.03 \pm     0.15 $  \\
 PGMS-77 & $     41.6 \pm      6.4 $ & $    -3.05 \pm     0.27 $ & $     33.8 \pm      3.6 $ & $    -2.47 \pm     0.16 $ & $     38.4 \pm      4.5 $ & $    -2.75 \pm     0.19 $  \\
 PGMS-82 & $     16.2 \pm      3.3 $ & $    -2.36 \pm     0.35 $ & $     18.0 \pm      3.0 $ & $    -2.51 \pm     0.27 $ & $     21.7 \pm      2.8 $ & $    -2.57 \pm     0.21 $  \\
 PGMS-87 & $     26.9 \pm      1.6 $ & $    -2.21 \pm     0.07 $ & $     11.3 \pm      0.7 $ & $    -1.74 \pm     0.07 $ & $     19.9 \pm      1.4 $ & $    -2.08 \pm     0.09 $  \\
\hline
\end{tabular}
\end{table*}
In most fields the spectra follow a power law behaviour that flattens at the high multipole end 
because of the noise contribution.  Exceptions are the four fields closest to 
the Galactic plane (PGMS-17 to PGMS-02) where a  power law modulated
by the beam window function dominates everywhere. .

We fit the angular power spectra to a power law modulated by the beam window function $W_\ell^B$
for the synchrotron component and a constant term $N$ for the noise:
\begin{equation}
  C_\ell^X = \left[ C_{200}^X \left(\frac{\ell}{200} \right)^{\beta^X}W_\ell^B + N \right] \,W_\ell^P\,,
\end{equation}
where 
$C_{200}^X$ is the spectrum at $\ell=200$ 
and $X = E, B, (E+B)/2$ denoting  $E$--Mode, $B$--Mode, and their mean 
($E$+$B$)/2. Possible residual contributions by point sources are accounted for by 
the constant term. 

Plots of the best fits are shown in Fig.~\ref{all_meas_specs:Fig} and~\ref{specs_E_B_EB:Fig}, 
while the parameters of the synchrotron component are reported in Table~\ref{bestFit:Tab}.

\begin{figure}
\centering
  \includegraphics[angle=90, width=1.0\hsize]{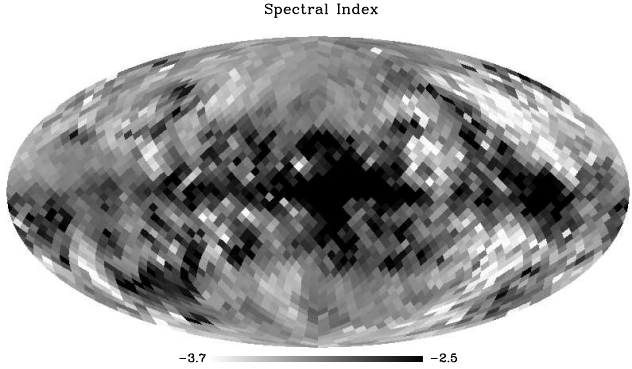}
\caption{ Map of frequency spectral index of the polarized synchrotron emission 
                 computed using the all-sky polarization maps at 1.4~GHz \citep{wolleben06,testori08}
                 and at 22.8~GHz (WMAP-5yr, \citealt{hinshaw09})\label{indexMap:Fig}
}
\end{figure}

\begin{figure}
\centering
  \includegraphics[angle=90, width=1.0\hsize]{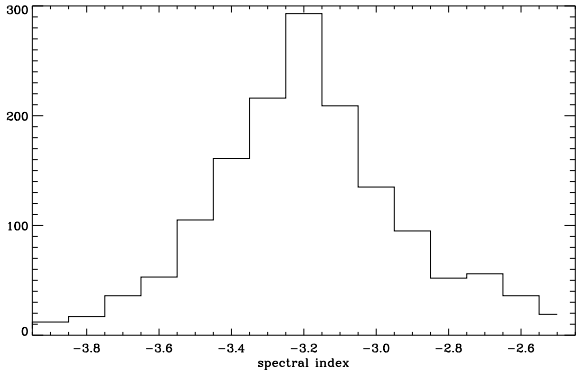}
\caption{Distribution of the spectral indexes reported in Fig.~\ref{indexMap:Fig}.
\label{specIndex:Fig}
}
\end{figure}

\begin{figure}
\centering
  \includegraphics[angle=0, width=1.0\hsize]{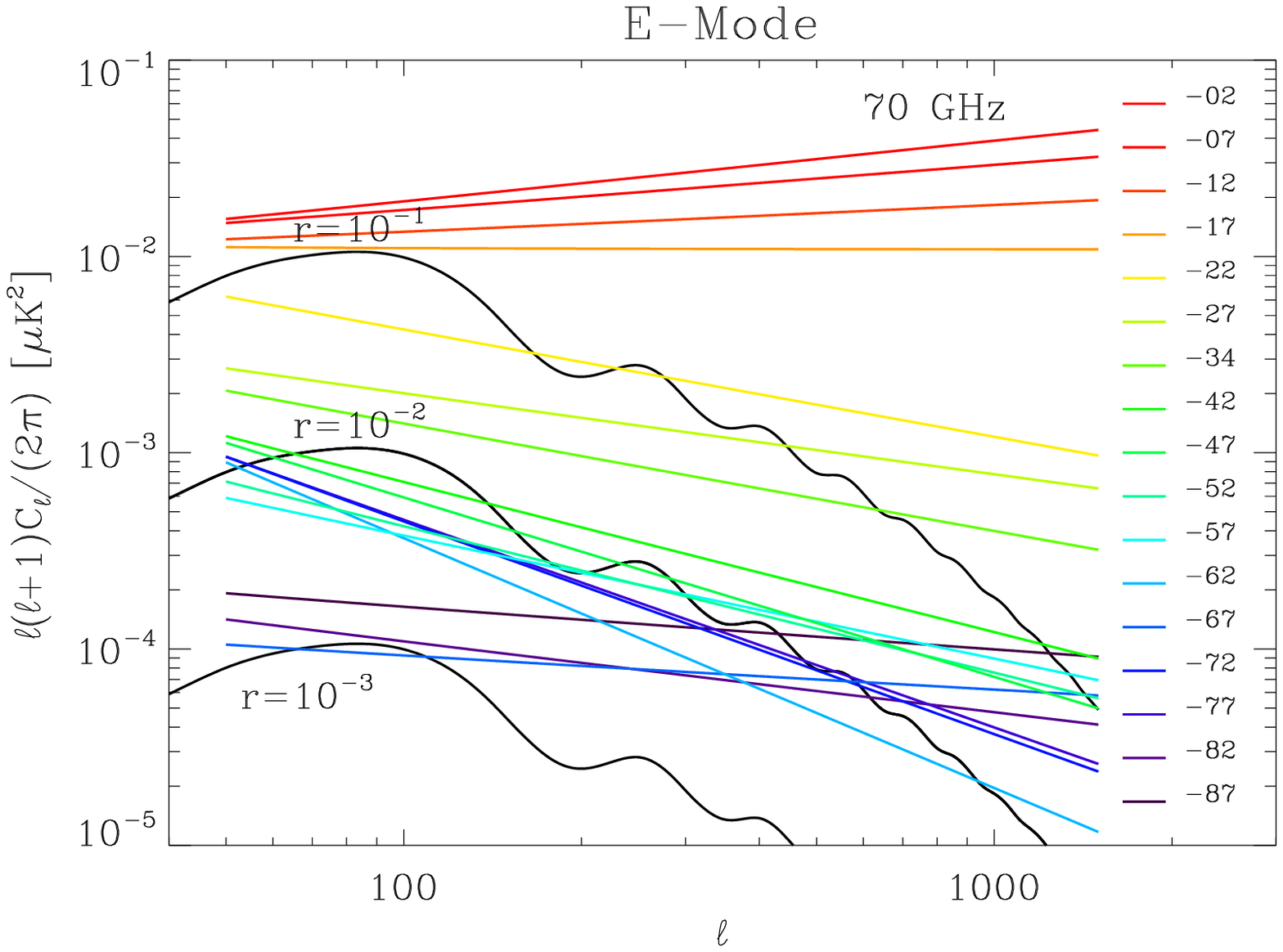}
  \includegraphics[angle=0, width=1.0\hsize]{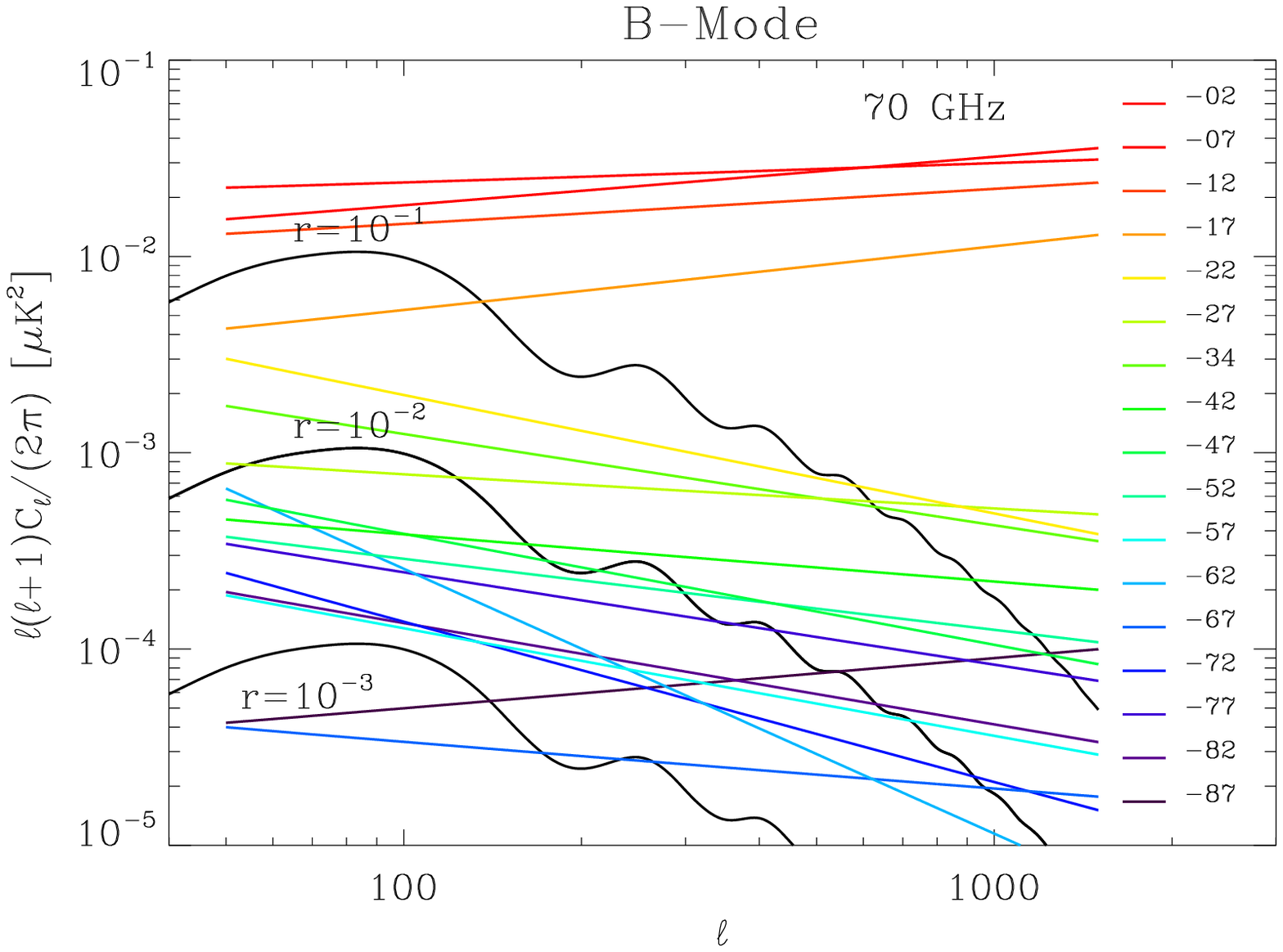}
  \includegraphics[angle=0, width=1.0\hsize]{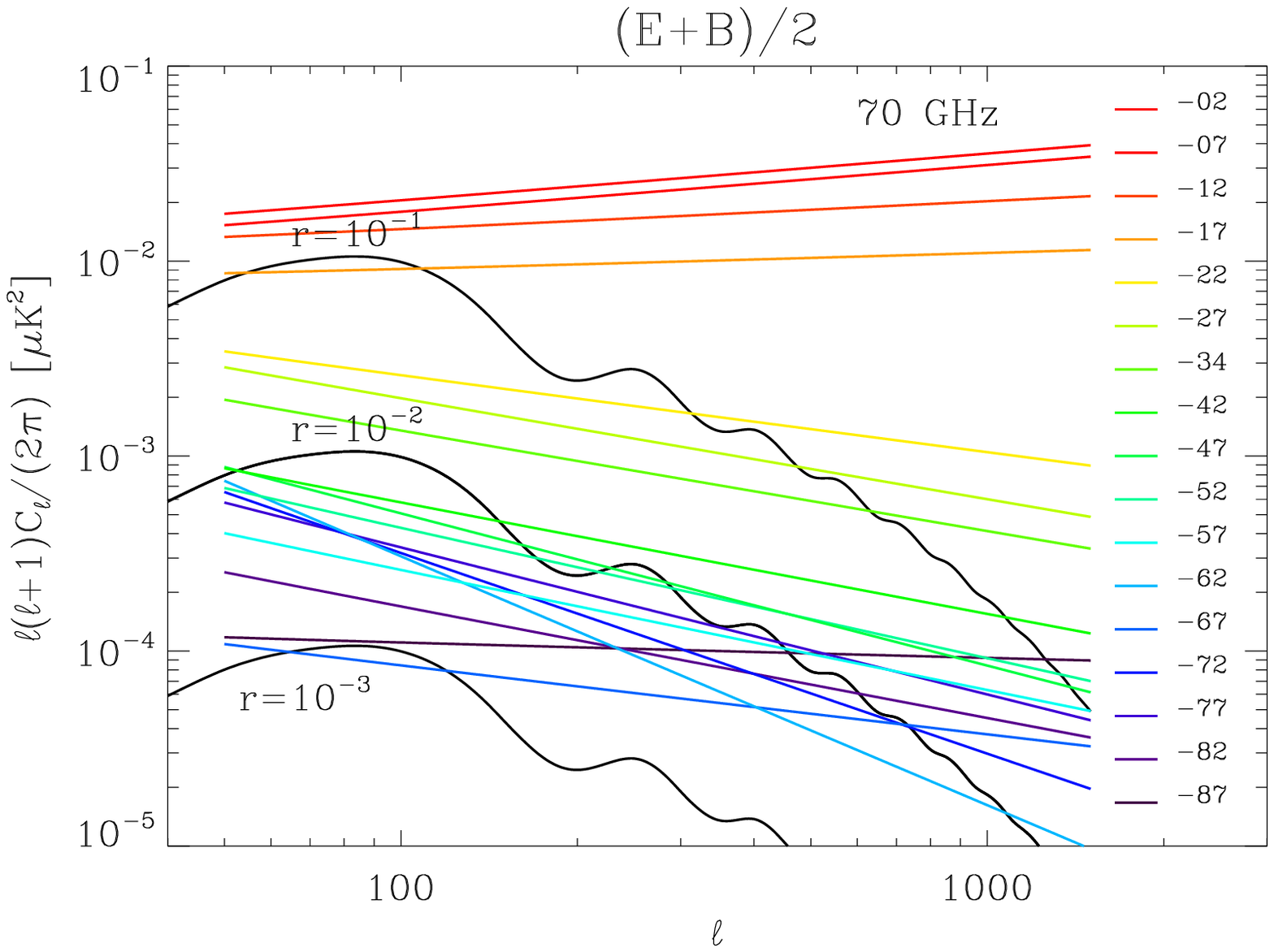}
\caption{ Best fits to the spectra of $E$--Mode (top),
                 $B$--Mode (mid), and their mean ($E$+$B$)/2 (bottom) of all PGMS fields. 
                 The plot reports only the power law component
                 which describes the synchrotron emission. All PGMS fields are plotted together
                 for a direct comparison of the behaviour with the Galactic latitude. The colour code goes from blue
                 throughout red by a rainbow palette for the areas from the south Galactic pole (PGMS-87) 
                 throughout  the Galactic plane (PGMS-02), respectively. The spectra are scaled to 70~GHz 
                 for a comparison with the CMB signal (frequency spectral index $\alpha = -3.1$). 
                 CMB $B$--Mode spectra for different values of tensor-to-scalar power ratio $r$ are shown
                 for comparison. The quantity plotted here is $\ell(\ell+1)/(2\pi) * C^X_\ell$, which provides
                 a direct estimate of the power distribution through the angular scales: a flat spectrum means
                 the power is evenly distributed; a decreasing spectrum (steep) is dominated by the largest scales; 
                 a rising spectrum (inverted) is dominated by the smallest scales. 
                \label{allSpec70:Fig}
}
\end{figure}

To analyse the behaviour of the synchrotron component, 
the power law component of
the best fit spectra are plotted together in Fig.~\ref{allSpec70:Fig} 
where they are also extrapolated to 70~GHz for comparison with the CMB signal. 
To determine the spectral index for the frequency extrapolation 
we computed a map of spectral index of the polarized synchrotron emission (Fig.~\ref{indexMap:Fig})
using the all-sky polarization surveys at 1.4~GHz \citep{wolleben06,testori08}
and 22.8 GHz (WMAP, \citealt{hinshaw09}). 
The index distribution at high Galactic latitudes ($|b|>30^\circ$) peaks at $\alpha_{\rm pol} = -3.21$ 
with dispersion $\Delta\alpha_{\rm pol} = 0.15$ (Fig.~\ref{specIndex:Fig}). 
This is consistent with the analysis of the WMAP-5yr data by \citet{gold09}, who find 
a polarized synchrotron spectral index  of $\alpha_{\rm WMAP} = -3.1$ 
in the WMAP frequency range,  and that of \citet{bernardi04} of total intensity data 
($\alpha_I = -3.1$ in the range 1.4--22.8 GHz).
We therefore assume a spectral index of $\alpha_{\rm synch} =-3.1$ for 
extrapolations up to the CMB frequency window. This assumption is 
somewhat  conservative and gives some margin to our conclusions.

\begin{figure}
\centering
  \includegraphics[angle=0, width=1.0\hsize]{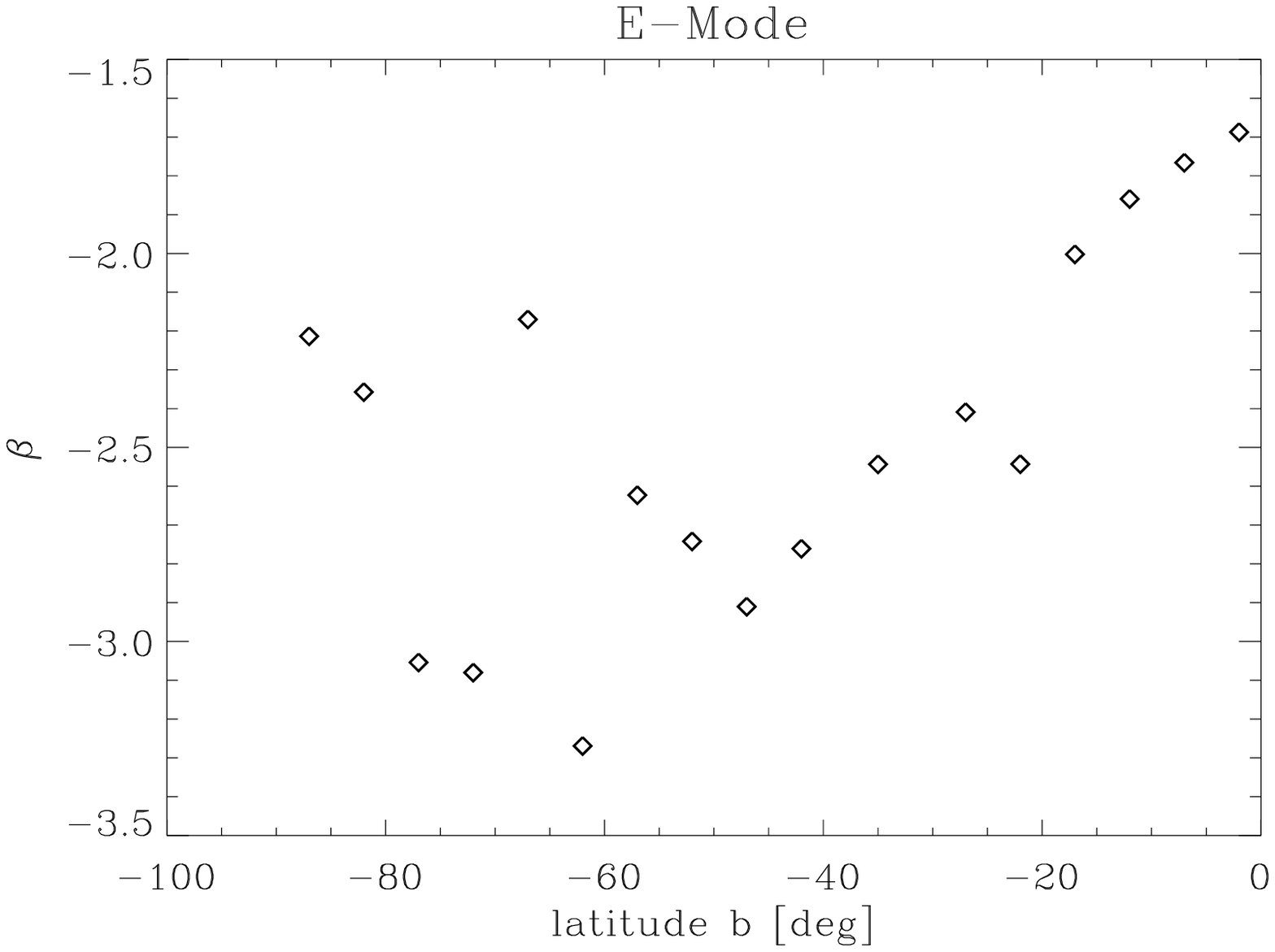}
  \includegraphics[angle=0, width=1.0\hsize]{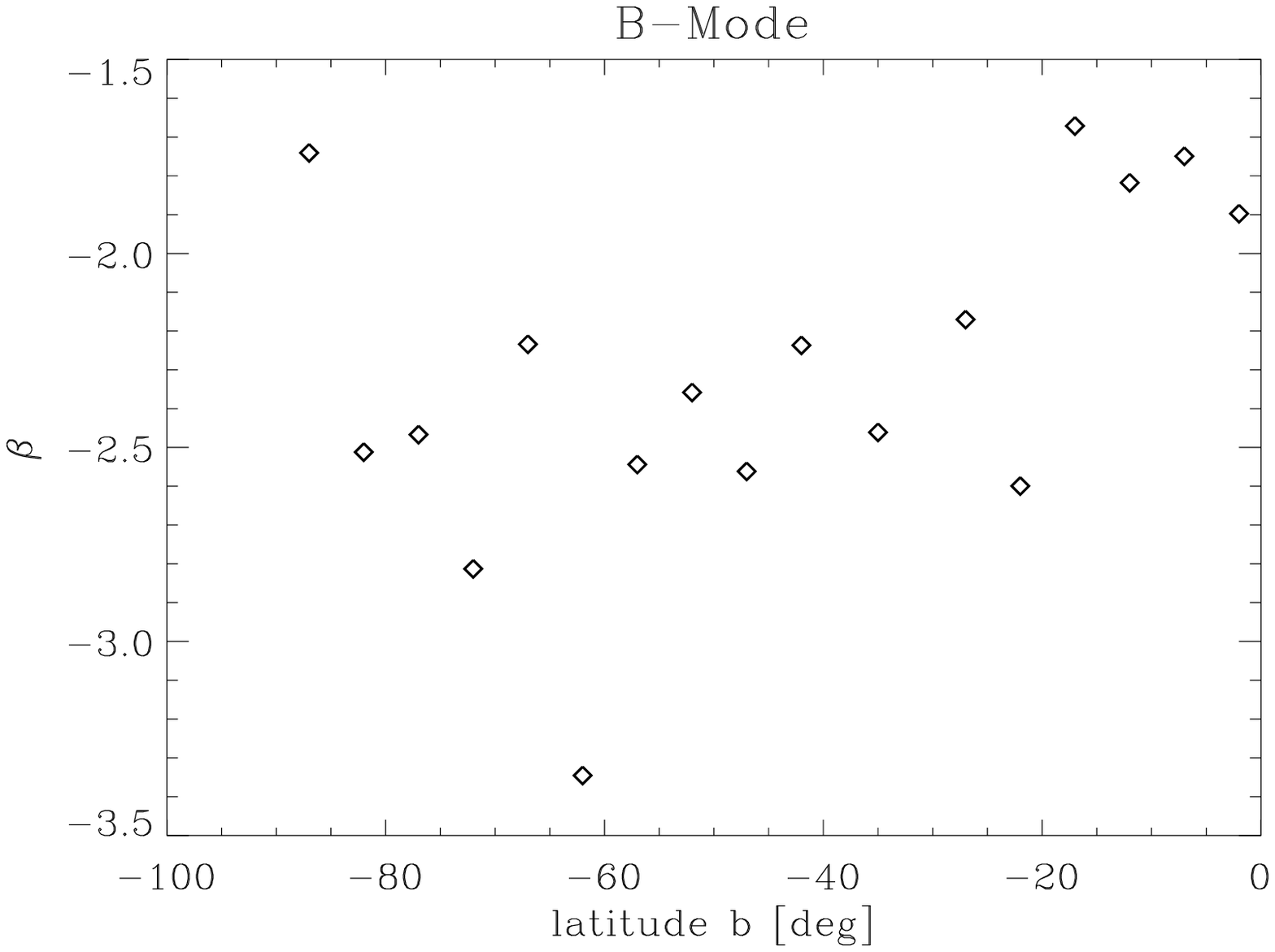}
  \includegraphics[angle=0, width=1.0\hsize]{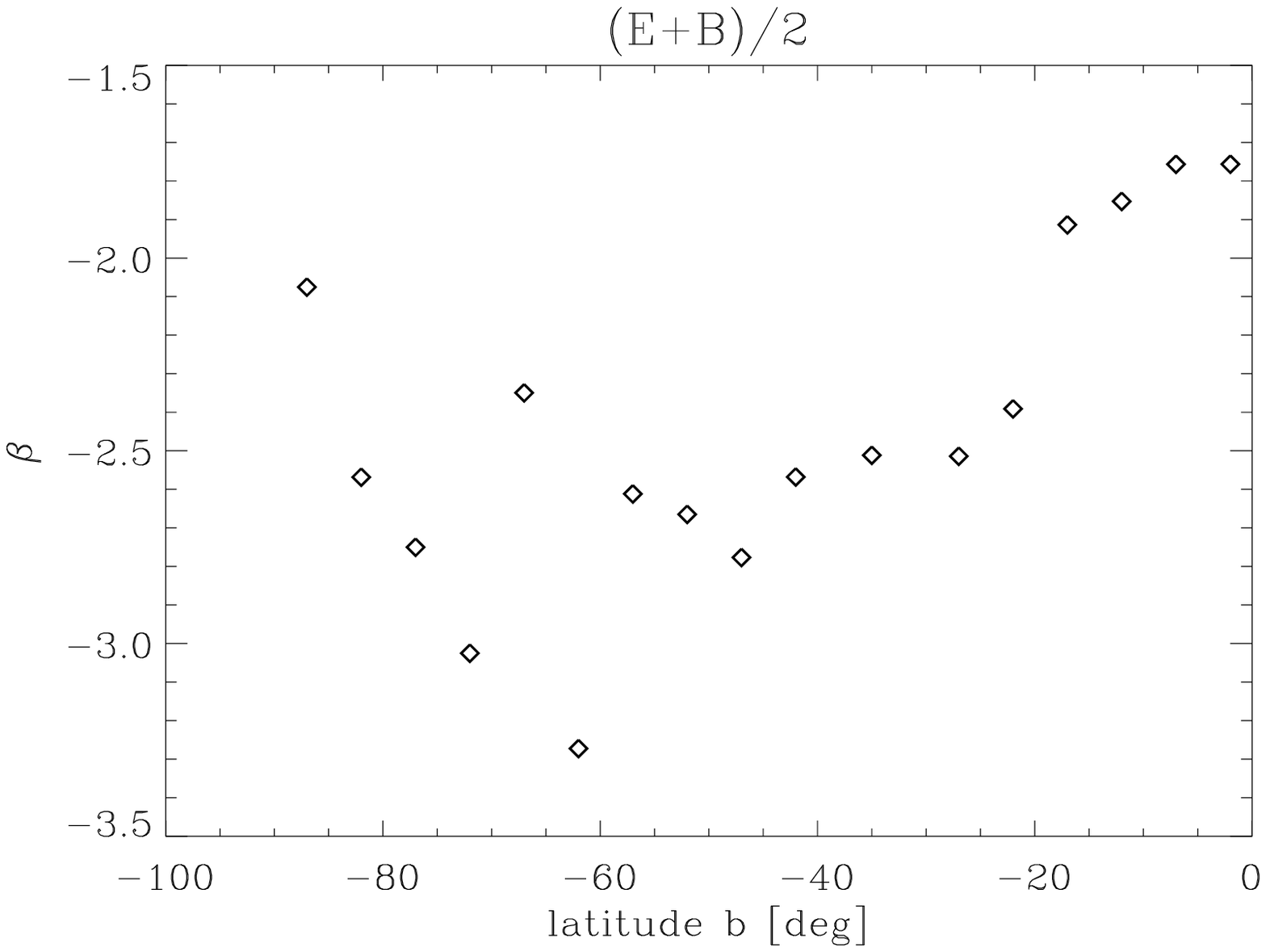}
\caption{Angular spectral slopes of the PGMS fields plotted against the field's centre latitude for 
                $E$--Mode (top), $B$--Mode (mid), and ($E$+$B$)/2.
                \label{slope:Fig}
}
\end{figure}

Distinguishing fields in terms of their spectral slopes, Fig.~\ref{allSpec70:Fig}  and Fig.~\ref{slope:Fig}
show the presence of two well defined regions:
\begin{itemize}
   \item{} The high and mid latitude fields ($b$~=~[-90$\degr$,~-20$\degr$]), with
               steep spectra ($\beta < -2.0$); slopes are distributed over the wide range
               $\beta\sim$~[-3.0,~-2.0] (except for  a few outliers). 
               There is no clear trend with latitude and the slopes 
               are rather uniformly distributed.  The median\footnote{We prefer to use the median to estimate the typical
                                                      angular slope in this region because of the possible significant 
                                                      deviations by the outliers.}
               is $\beta_{\rm med}^{(E+B)/2} = -2.6$ (Table~\ref{slope_mh:Tab}). 
               The dispersion $\sigma_{\beta^{(E+B)/2}}=0.24$ 
               is significantly larger than the individual 
               measurement errors, meaning that this wide spread is 
               an intrinsic property of the synchrotron emission at these latitudes.
   \item{} The low latitude fields ($b=[-20\degr,~0\degr]$), which show 
                inverted spectra ($\beta > -2.0$); the slopes lie in a much 
                narrower range  ($\beta^{(E+B)/2} = [-1.90,-1.75]$)
                with dispersion $\sigma_{\beta^{(E+B)/2}}=0.08$ (Table~\ref{slope_l:Tab}).
                All spectra have mean and median slope $\bar{\beta}^X = \beta^X_{\rm med} = -1.8$, 
                which can be considered the typical value of this region.
\end{itemize}

This change from steep to inverted spectra is quite sudden and 
clearly separates two different environments:
the mid-high latitudes, characterised by a smooth emission with most of the power
on large angular scales,
and the disc fields, whose power is more evenly distributed
with a slight predominance of the small scales. 
\begin{table}
\label{slope_mh:Tab}
\centering
\caption{Mean ($\bar{\beta}^X$), median ($\beta^X_{\rm med}$), and dispersion ($\sigma_\beta$) 
               of the angular slopes  at mid-high Galactic latitudes for all the three spectra $X=E,B,(E+B)/2$.}
\begin{tabular}{@{}c|ccc@{}}
\hline
         \multicolumn{4}{c}{MID-HIGH latitudes} \\
\hline
X  &  $\bar{\beta}^X$  &   $\beta^X_{\rm med}$  &  $\sigma_{\beta^X}$  \\
\hline
$E$   &  -2.67   &    -2.62    &  0.34     \\
$B$   &  -2.39   &    -2.46    &  0.27     \\
$(E+B)/2$   &  -2.57   &    -2.57    &  0.24    \\
\hline
\end{tabular}
\end{table}
\begin{table}
\centering
\caption{As for Table~\ref{slope_mh:Tab} but for low Galactic latitudes.}
\begin{tabular}{@{}c|ccc@{}}
\hline
         \multicolumn{4}{c}{LOW latitudes} \\
\hline
X  &  $\bar{\beta}^X$  &   $\beta^X_{\rm med}$  &  $\sigma_{\beta^X}$  \\
\hline
$E$   &  -1.83   &    -1.81    &  0.14     \\
$B$   &  -1.78   &    -1.78    &  0.10     \\
$(E+B)/2$   &  -1.82   &    -1.80    &  0.08    \\
\hline
\label{slope_l:Tab}
\end{tabular}
\end{table}

Does this change indicate an intrinsic feature of the polarized emission of the disc, or is it the  
effect of Faraday modulation, which transfers power
from large to small angular scales in the low latitude fields?
The answer is unclear with the information available, but some points can be noted.
In the disc the ISM is more turbulent than in the halo and the intrinsic emission
might have more power on small angular scales.
In addition, the low-latitude lines of sight go through much more ISM including more 
distant structures; these are expected to give more power to the small angular scales.
Also the smooth emission of the two highest latitude disc fields 
(PGMS-12 and PGMS-17) make the presence of significant Faraday depolarization unlikely.
Finally, we have computed the power spectrum of the individual frequency channels
to search for a possible variation of the angular slope with frequency. 
Since the lowest frequencies would be more affected, steeper spectra at highest 
frequencies would support the presence of FR effects. We find that all the four disc
fields have non-significant slope variation compatible with zero within 
1.0--1.5 sigma, with the only exception of PGMS-02 which 
approaches 2-sigma.
All these points support the view that the structure of the
low-latitude polarized emission derives from the intrinsic nature of
the synchrotron emitting regions close to the plane, and is not
imposed by Faraday depolarization along the propagation path.
\begin{figure}
\centering
  \includegraphics[angle=0, width=1.0\hsize]{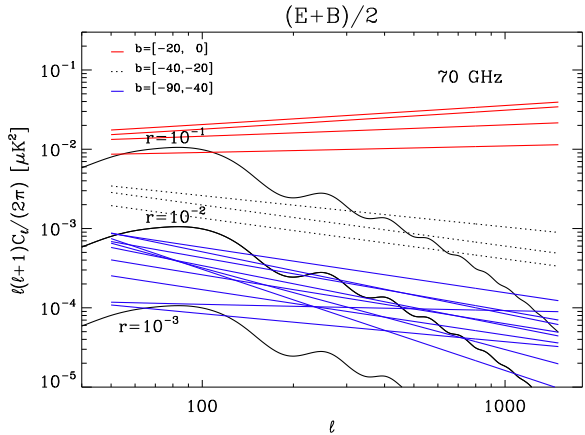}
\caption{ As bottom panel of  Fig.~\ref{allSpec70:Fig}, but with spectra grouped 
  according to latitude region.  The disc fields (solid, red: $b = [-20\degr, 0\degr]$) 
  are  clearly
  distinguished by their spectral slopes and higher amplitudes, and the fields in the
  transition region (dotted, black: $b = [-40\degr,-20\degr]$) have amplitudes quite distinct from both the disc
  and halo fields (solid, blue: $b = [-90\degr,-40\degr]$). 
  \label{allSpec70_3col:Fig}
}
\end{figure}

Considering the amplitude distribution of the PGMS fields, we further
divide the mid-high latitude region identified above 
into an halo 
($b = [-90\degr,-40\degr]$) and transition region ($b =
[-40\degr,-20\degr]$).  Thus we identify three distinct latitude
sections:  two main regions (disc and high latitudes) well distinguished by both emission power
 and structure of the emission, and an extended transition  
 about $20\degr$ wide connecting them.
 Fig.~\ref{allSpec70_3col:Fig} reports the spectra for all field,
 showing how they belong to the three regions, which:
\begin{itemize}
   \item {} Halo  ($b = [-90\degr,-40\degr]$):  the emission is weak here and, scaled to 70~GHz,
             is between the peaks of CMB models with $r=10^{-3}$ and $r=10^{-2}$. 
             The weakest fields (PGMS-87 and PGMS-67) match models with 
             $r=10^{-3}$. The fluctuations
              from field-to-field dominate with no clear trend with latitude. A weak
              trend might be present with the emission power increasing toward lower latitudes, 
              but  the effect is a minor  in comparison to the dominant field-to-field fluctuations.
   \item {} Galactic disc  ($b = [-20\degr,0\degr]$): the emission is stronger, about two orders
      of magnitude brighter than that of the halo. Within the area there is
   no large variation of the emission power, but slight increase toward the Galactic plane is evident.
   \item {}  Transition strip  ($b = [-40\degr,-20\degr]$): here a transition between the faint high latitudes 
   and the bright disc occurs. This is clearer at large scales, where the northernmost 
   field (PGMS-22) is almost as bright as the weakest disc field (PGMS-17) and the southermost
   field approaches the upper end of the halo brightness range. 
\end{itemize}

An important consequence is the identification of a clear transition between disc and halo.
The sudden change in the angular spectral slope at $|b|\sim 20\degr$ and the approximately constant  
emission power from the Galactic plane up to that transition clearly separate 
the $20\degr$ equatorial zone from the higher latitudes. Characterised by a more complex structure of the ISM
this area can be associated with the Galactic disc.

A second environment characterised by steep spectra and low emission is clearly present for $|b| > 40\degr$. 
Both angular slope and amplitude exhibit wide fluctuations without any
clear trend with latitude.  We consider this high Galactic latitude
section as a single environment with regard to its
polarized synchrotron emission properties. Characterised by a smoother emission 
and simpler ISM structure, this area we  associate with the Galactic halo.
\begin{table*}
\centering
\caption{Best fit amplitude $C^X_\ell$ and angular spectral slope $\beta^X$ 
of the mean spectrum of the PGMS halo section ($X=E, B, (E+B)/2$ denotes $E$--, $B$--Mode, and ($E$+$B$)/2,
respectively).
\label{bestFit_halo:Tab} 
}
\begin{tabular}{@{}lcccccc@{}}
\hline
 Field  &  $C^E_{200}$~[$\mu$K$^2$]   &  $\beta^E$  &  
           $C^B_{200}$~[$\mu$K$^2$]   &  $\beta^B$  &  
           $C^{(E+B)/2}_{200}$~[$\mu$K$^2$]  &  $\beta^{(E+B)/2}$  \\ 
\hline
 PGMS HALO & $     40.7 \pm      3.2 $ & $    -2.65 \pm     0.10 $ & $     28.5 \pm      3.0 $ & $    -2.64 \pm     0.12 $ & $     35.5 \pm      2.9 $ & $    -2.60 \pm     0.09 $  \\
\hline
\end{tabular}
\end{table*}
\begin{figure}
\centering
  \includegraphics[angle=0, width=1.0\hsize]{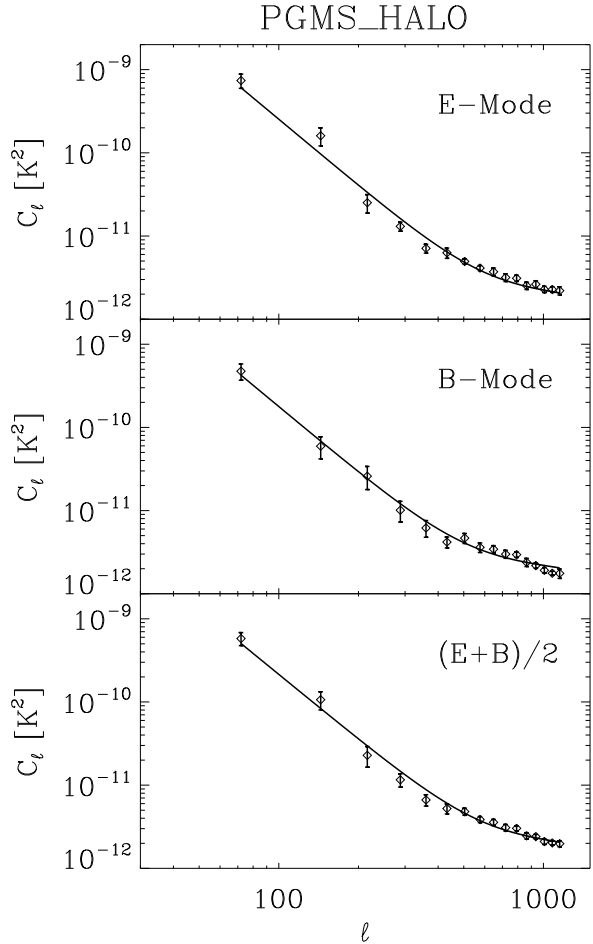}
\caption{Mean of the power spectra of the halo fields ($b=[-40\degr, -90\degr]$)
                for $E$--Mode (top), $B$--Mode (mid), and their mean ($E$+$B$)/2 (bottom).
                Both the mean of the measured spectra (diamonds) and 
                its best fit curve (solid) are plotted.
                \label{specs_halo_E_B_EB:Fig}
}
\end{figure}
\begin{figure}
\centering
  \includegraphics[angle=0, width=1.0\hsize]{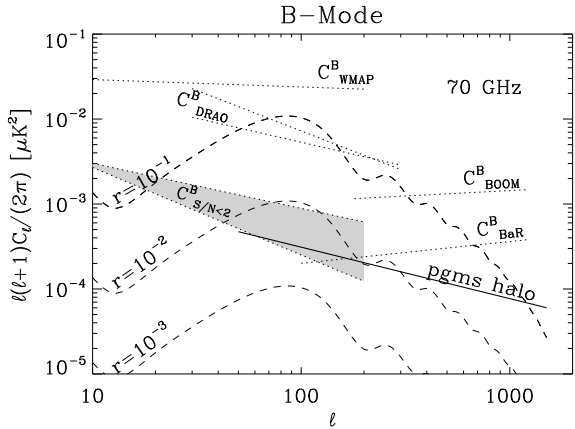}
\caption{Best fit of the mean halo spectrum $C^{(E+B)/2}_\ell$ scaled to 70~GHz (solid)
                alongside CMB $B$--Mode spectra for different $r$ values (dashed). 
                The mean synchrotron contamination at all high Galactic latitudes in also shown
                for comparison ($C^B_{\rm WMAP}$)
                together with the previous estimates in other low emission regions:
                 $C^B_{S/N < 2}$ (shaded area),  $C_{\rm DRAO}$, $C^B_{\rm BOOM}$, and 
                 $C^B_{\rm BaR}$ (see Fig.~\ref{allSpec70:Fig} for details).
                \label{specs_halo_70:Fig}
}
\end{figure}

The emission of the halo section is very weak. 
In spite of large fluctuations, once scaled to 70~GHz the synchrotron component 
is equivalent to $r$ values between $1\times10^{-3}$ and $7\times10^{-3}$, which
matches the weakest areas observed so far in polarization. It is worth noticing that 
PGMS fields PGMS-87 and PGMS-67 have the weakest polarized synchrotron
emission observed so far.

The high Galactic latitudes above 40$\degr$ are thus just one 
environment, at least in a low emission region not contaminated by local anomalies like
the area intersected by PGMS. This is very important for CMB investigations, since it tells
that, in principle, it is possible to find large areas with optimal conditions
(extended over $50\degr$, in the PGMS case).

It is thus important to measure the mean emission properties of 
the entire halo section ($|b| > 40\degr$).
The mean spectrum of the 10 halo fields and its best fit are plotted in
Fig.~\ref{specs_halo_E_B_EB:Fig}; Table~\ref{bestFit_halo:Tab} gives
the best fit parameters; the extrapolation to 70~GHz is shown in
Fig.~\ref {specs_halo_70:Fig}.

The angular slope is $\beta^X \sim -2.6$ for all the three spectra $E$, $B$, and ($E$+$B$)/2 
and is thus be considered the typical slope of the halo section.  Note
that this matches the slope measured at high latitudes by WMAP at 22.8~GHz
\citep{page07}, indicating  that the power distribution through the angular scales
is the same at 2.3 and 22.8 GHz.  This further argues against the
significance of Faraday depolarization, which would have transferred
power from large to small scales and made the angular spectra
frequency dependent.

Once scaled to 70~GHz, the amplitude is equivalent to 
\begin{equation}
r_{\rm halo} = (3.3\pm0.4)\times10^{-3}, 
\end{equation}
roughly in the middle of the range covered by the individual fields. 
As mentioned earlier,  this is a very low level and corresponds to 
the faintest areas observed previously, which thus seem to
be more the normal condition of the low emission regions 
rather than {\it lucky} exceptions. 

Finally, the field PGMS-34 deserves mention as the area  
observed by the experiment BOOMERanG.  It lies in the transition region, and  
although at the high-latitude end of this zone, has emission about  
five times greater than fields in the halo.  It is suitable for  
measuring the stronger $E$-mode (the aim of the 2003 BOOMERanG flight),
but our results identify more suitable fields for detecting the B-mode.

\section{Dust emission in the PGMS halo section}
\label{dust:Sect}

The dust emission is the other most significant foreground for CMB observations. It 
has a positive frequency spectral index and  dominates the foreground budget
 at high frequency.  
 
An estimate of the local  dust contribution  in the same portion of halo covered by the PGMS
is thus important to understand the overall limits of a CMB $B$--Mode detection in 
that area. 

\begin{figure}
\centering
  \includegraphics[angle=0, width=1.0\hsize]{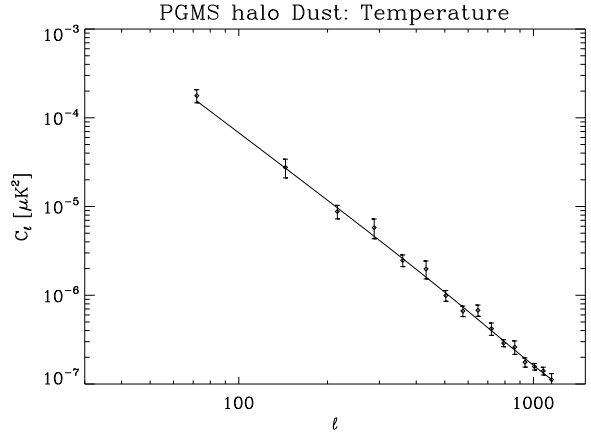}
\caption{ Temperature power spectrum of the dust emission model at 
                94~GHz in the halo section of the PGMS ($b=[-90\degr,-40\deg]$). 
                The spectrum is the mean of those computed in the individual PGMS fields.
                Both the mean of the computed spectra (diamonds) and its best fit curve (solid) 
                are plotted. The latter is a power law with slope $\beta^T_d = -2.50 \pm     0.14$ 
                 (see text and Table~\ref{bestFit_halo_dust:Tab}).
                \label{specs_halo_dust:Fig}
}
\end{figure}

However,  no polarized dust emission has been detected over the PGMS region, and
even the total intensity dust map of WMAP is noise dominated in that area.
We therefore use the \citet*{fi99} model of the total intensity 
dust emission applying an assumed polarization fraction. 
First, we generate maps at 94~GHz using their model-8 for each of the 
10~PGMS fields at $|b|>40\degr$. The temperature 
power spectra of each are then computed and averaged together to estimate the mean 
conditions of the whole $5\degr \times 50\degr$ section. 
Fig.~\ref{specs_halo_dust:Fig} shows both the mean spectrum and its best fit,
whose parameters are reported in Table~\ref{bestFit_halo_dust:Tab}. 
Within the errors, the angular slope matches well that of the synchrotron. 
\begin{table}
\centering
\caption{Best fit amplitude $C^T_{200}$ and angular spectral slope $\beta^T_d$ 
of the mean Temperature dust spectrum at 94 GHz in the halo section 
of the PGMS ($|b|>40\degr$). 
\label{bestFit_halo_dust:Tab} 
}
\begin{tabular}{@{}ccc@{}}
\hline
 Field  &  $C^T_{200}$~[$\mu$K$^2$]   &  $\beta^T_d$  \\ 
\hline
 PGMS halo Dust & $ (12.0  \pm     1.4)\times 10^{-6} $ & $    -2.50 \pm     0.14 $  \\
\hline
\end{tabular}
\end{table}

The total polarized spectrum is estimated from this temperature spectrum 
assuming a polarization fraction $f_{\rm pol} = 0.10$, as inferred for 
high Galactic latitudes from the Archeops experiment data \citep{be04,ponthieu05}. 
The spectrum is further divided by two to account for an even
distribution of power between $E$-- and $B$--Mode, a
reasonable assumption for the Galactic emission.

For frequency extrapolations, we use a single Planck function
modulated by a power law $T_d \propto \nu^{\alpha_d}/(e^{h\nu \over kT} -1)$
with index $\alpha_d = 2.67$ and temperature $T = 16.2$~K, 
which  reproduces the \citet{fi99} model well in the range $[70,150]$~GHz. 
It is worth noticing that at 94~GHz that function is well approximated by a
power law with index $\alpha_d^{\rm PL} = 1.5$, consistent
with the 5-yr WMAP result of $\alpha_d^{\rm WMAP} = 1.8 \pm 0.3 \pm0.2$
(statistical and systematic error, \citealt{gold09}).

\begin{figure}
\centering
  \includegraphics[angle=0, width=1.0\hsize]{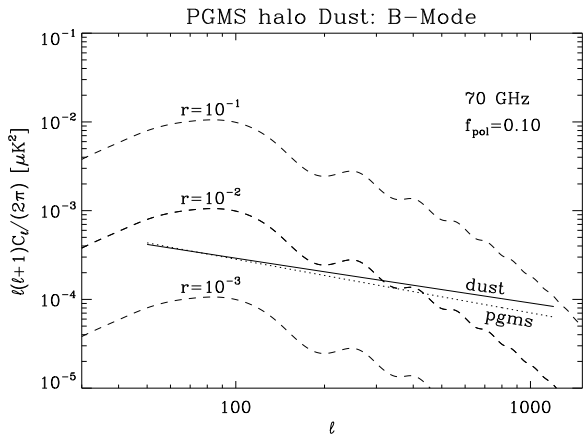}
\caption{ Estimates of the dust polarized emission in the halo section of the PGMS at 70 GHz (solid).
                 The best fit to the temperature spectrum is used assuming a dust polarization fraction
                 $f_{\rm pol} = 0.10$ and then scaled to 70~GHz.
                 The synchrotron emission spectrum estimated in Section~\ref{aps:Sect} is shown for 
                 comparison (dotted), as well as the  CMB $B$--Mode  spectra for three different $r$ values.
                \label{specs_halo_dust_70:Fig}
}
\end{figure}

Our estimate of the $B$--Mode polarized dust spectrum at 70~GHz is given
in Fig.~\ref{specs_halo_dust_70:Fig}. At this frequency the two  
components of Galactic polarized emission
are approximately equal, and so the  
total polarized foregound is at a minimum.
This frequency is mildly dependent on assumed polarized fraction of  
the dust component: alternate assumptions of five percent or  
twenty percent shift the frequency of minimum to 80~GHz and 60~GHz  
respectively.

This result is similar to that of the general high Galactic latitudes 
and suggests that the synchrotron-to-dust power ratio is only marginally dependent
on the strength of the Galactic emission. 

\section{Limits on \lowercase{$\bmath{r}$} }
\label{limits:Sect}

\begin{figure*}
\centering
  \includegraphics[angle=0, width=0.495\hsize]{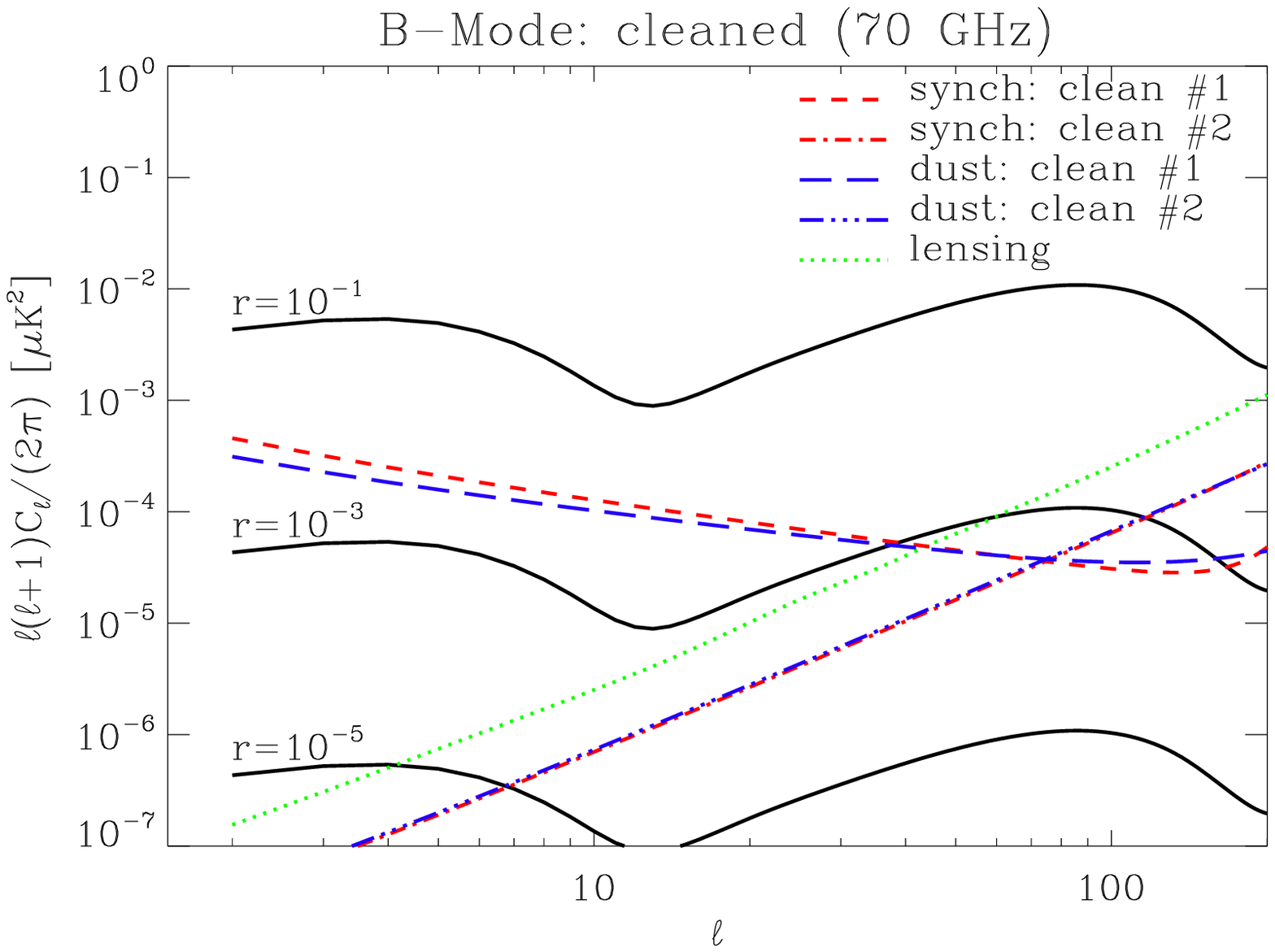}
  \includegraphics[angle=0, width=0.495\hsize]{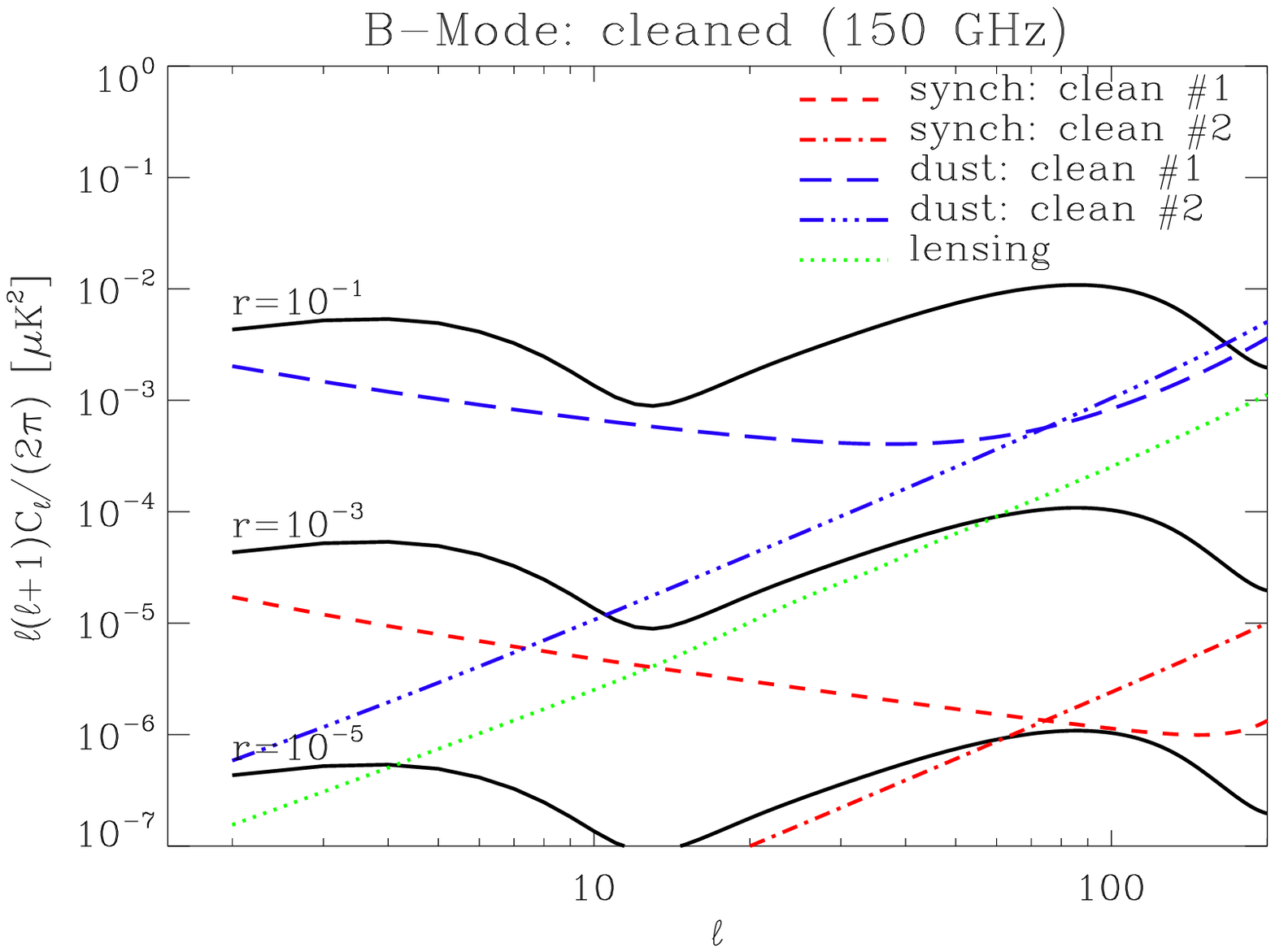}
\caption{Residual contaminations after foreground cleaning in the PGMS halo section at 70 (left) and 150~GHz (right) 
                for synchrotron (red) and dust emission (blue). Results for both the two cleaning methods
                described in text are shown. The gravitational lensing contribution is also shown 
                assuming that cleaning has reduced its contribution by a factor 10 in the power spectrum (green).
 \label{spec_halo_clean:Fig}
}
\end{figure*}

\begin{table*}
 \centering 
  \caption{Detection limits of $r$ at 3-sigma C.L. ($\delta r$) at 70 and 150~GHz in the PGMS halo area
                  and in a 2500-deg$^2$ region assumed to have equivalent foreground contamination
                  (see text). 
                  Results for both cleaning types described in the text are reported.}
  \begin{tabular}{@{}cccc@{}}
  \hline
  Area                   &  clean type & $\delta r$ (70 GHz)  & $\delta r$ (150 GHz) \\
   \hline
  PGMS                        &  clean \#1 &  $2.2\times10^{-3}$    &    $8.0\times10^{-3}$        \\
  PGMS                        &  clean \#2 &  $1.5\times10^{-3}$    &    $5.5\times10^{-3}$        \\
  \hline
  2500-deg$^2$         &  clean \#1 &    $1.2\times10^{-3}$    &     $4.2\times10^{-3}$     \\
  2500-deg$^2$         &  clean \#2 &    $0.5\times10^{-3}$    &     $1.8\times10^{-3}$     \\
   \hline
  \end{tabular}
 \label{delta_r:Tab}
\end{table*}

To estimate the detection limits of $r$ in the presence of the
foreground contamination in the PGMS halo section,
we consider an experiment with resolution $\theta_{\rm cmb} = 1^\circ$ 
at the CMB frequency channel to have adequate sensitivity at the $\ell \sim 90$ peak.

We also account for the cleaning provided by foreground separation 
techniques. We consider the two cases discussed
by \citet{tucci05}: 
\begin {enumerate}
  \item{}  cleaning by subtracting the foreground map scaled to higher frequencies
               using just one frequency spectral index for all pixels. It is a coarse method and
               represents a worst case.  The residual contamination depends on the 
               spread of spectral indices in the area.
               We use $\Delta\alpha_s = 0.15$  for the synchrotron component
               (Section~\ref{aps:Sect}; see also \citealt{bernardi04,gold09}), 
               and assume the same dispersion for the dust emission ($\Delta\alpha_d = 0.15$). 
               We refer to this method as {\it clean \#1}.
  \item{} cleaning by foreground subtraction, but assuming knowledge of the frequency spectral index for any individual pixel.
              In this case the amplitude of the residual contamination 
              depends on the measurement error of the frequency slopes.
              Here we assume a combination of the PGMS data with a
              {\it synchrotron channel} at 22 GHz onboard the CMB experiment 
              with resolution $\theta_{\rm s} = 1^\circ$ and sufficient sensitivity to give 
              $S/N=5$. 
             For the dust emission, we assume that the CMB experiment includes 
             a {\it dust channel} at 350~GHz with a sensitivity to allow
             a $S/N=5$ and resolution scaled from the CMB channel 
             ($\theta_{\rm d} = \theta_{\rm cmb}\nu_{\rm cmb}/\nu_{\rm d}$, where 
              $\nu_{\rm cmb}$ and $\nu_{\rm d}$ are the frequencies of the CMB and dust channel).
             We refer to this method as {\it clean \#2}.
\end {enumerate}
These two methods are at the two ends of cleaning capabilities (\#1 is coarser
and less efficient, \#2 is finer and more efficient), so give a good idea
of the range of possible performance.

The left panel of Fig.~\ref{spec_halo_clean:Fig} 
shows the residual contaminations left by these two methods at 
70~GHz in the PGMS halo section using the method of \citet{tucci05}. 
The effect of  method \#1 is to reduce the amplitude of the contamination
but preserve the shape of the power spectrum  (for instance, 
the synchrotron residual preserves the angular spectral index of -2.6). 
Method \#2 gives comparable results at the $\ell \sim 90$  
CMB peak, but has a flat white--noise--like spectrum which 
performs much better on large angular scales. 
While the simpler clean~\#1 looks appropriate as the target 
is the peak at $\ell \sim 90$, 
the clean~\#2 is better suited for the reionization 
bump at larger scales.

With such low levels of residual foreground contribution 
the effects of gravitational lensing become a dominant part of the 
contamination budget. Its subtraction is therefore also required 
to realise the benefit of the low Galactic emission of the PGMS strip.
Gravitational lensing effects can be cleaned using high resolution data (10~arcmin or better, 
\citealt{seljak04}) and here we assume that it can be reduced by
10~fold (\citealt{seljak04}).

The Fisher information matrix is used to estimate the detection limits
\citep*{tegmark97b,tegmark97a}.
If $r$ is the only parameter to be measured, 
(other cosmological parameters being provided by Temperature 
and $E$--Mode spectrum from other experiments such as
WMAP or  PLANCK), the Fisher matrix reduces to the scalar
\begin{equation}
   \mathcal{F}_{rr} = \sum_\ell \frac{1}{(\Delta C_{\ell}^B)^2}  \left(\frac{\partial C_\ell^B}{\partial r}\right)^2
\end{equation}
and the rms error on $r$ is 
\begin{equation}
  \sigma_r = \mathcal{F}_{rr}^{-1/2}. 
\end{equation}
The uncertainty $\Delta C_{\ell}^B$  of the $B$--mode spectrum
is a function of the CMB spectrum $C_{\ell}^{B,{\rm cmb}}$ and the cleaning residuals
of synchrotron, dust, and gravitational lensing:
\begin{eqnarray}
   \Delta C_{\ell}^B  &=& \sqrt{\frac{2}{(2\ell+1)\Delta\ell\,f_{\rm sky}}}\, C_{\ell}^{B,{\rm cmb}} \nonumber \\
   &+& \left(C_{\ell}^{B, {\rm synch-res}} + C_{\ell}^{B,{\rm dust-res}} + C_{\ell}^{B,{\rm lens-res}} \right),
\end{eqnarray}
where $\Delta\ell$ is the 
width of the multipole bins and $f_{\rm sky}$ is the sky
coverage fraction. 
As a set of cosmological parameters we use the best fits of the 
5-yr WMAP data release \citep{komatsu09}.

At 70~GHz the detection limits of $r$ in the PGMS halo region is 
$\delta r \sim 2\times 10^{-3}$ (3-sigma C.L.) for both cleaning methods
(Table~\ref{delta_r:Tab}). This is a very low level which
makes the PGMS strip an excellent target for CMB experiments and enables accessing levels 
of $r$ much lower than previously estimated for areas of comparable size
(e.g. \citealt{tucci05,verde06}, who used higher foreground levels estimated 
from total intensity data).
An important point is that there is only a marginal benefit from using
the more sophisticated cleaning method. For a 250-deg$^2$ area 
most of the sensitivity resides in the $\ell \sim 90$ peak, 
where the dominant residual is the gravitational lensing and a better cleaning
of the other contaminants is not critical.

This result is therefore quite robust since it is based on actual measurements 
of the foreground contamination in a specific area and 
is marginally dependent on the cleaning method. 
Moreover, the leading residual term is the gravitational lensing, giving a 
good margin against errors in our dust polarization fraction assumption. 

It is also important to estimate $\delta r$ for 150~GHz, a frequency that,
 although far from the foreground minimum, is  preferred 
by experiments based on bolometric detectors.
However, as shown in Fig.~\ref{spec_halo_clean:Fig}, right panel, 
the major residual at this frequency is dust emission, making the result
dependent on the assumed dust polarization fraction $f_{\rm pol}$. 
A value of $f_{\rm pol} = 0.10$  gives the limit $\delta r = 6$--$8\times 10^{-3}$ 
(Table~\ref{delta_r:Tab}), while
the goal of the most advanced sub-orbital experiments planned for 
the next years ($r \sim1\times 10^{-2}$) is still achievable
under the reasonable assumption that  $f_{\rm pol} < 0.12$. 
 
 The PGMS halo region is a narrow strip 50$^\circ$ long and it is unlikely that its optimal 
conditions are confined to its 5$^\circ$ width.
Therefore we consider that a larger area, say of  $50\degr \times
50\degr$ extent, could be identified with properties comparable to
those of the PGMS halo region.  Such an area, about 6 per cent of the
sky, matches in size
the southern portion of the low emission region identified 
in the WMAP data \citep{carretti06b}.

The detection limit achievable over such an area drops to 
 $\delta r=5\times 10^{-4}$ (3-sigma C.L.) if method \#2 is applied, or
$\delta r=1.2\times 10^{-3}$ under 
the coarser method \#1 (Table~\ref{delta_r:Tab}).  

Note that in this case there is a significant difference between
the two cleaning methods, justifying the use of the more
sophisticated method \#2. This is because the larger area makes measurement 
of the lowest multipole components relevant, where method~\#1 is not effective.

\section{Summary and Conclusions}
\label{conc:Sect}

The PGMS has mapped the radio polarized emission at 
all Galactic latitudes in a 5~degree strip at a frequency sufficiently
high not to be affected by Faraday depolarization
and with sufficient  sensitivity to detect the signal in low emission regions. 
It is the largest area observed so far at high Galactic latitude
uncontaminated by large local structures.
 
This has enabled us to investigate the behaviour of the polarized emission with latitude
by computing the polarized angular power spectrum in 17 fields from the Galactic plane
to the South Galactic pole. 
We can distinguish three latitude sections: two main regions well 
distinguished by both brightness and structure of the emission (disc and halo), and
an extended transition connecting them. In detail they are: 
\begin{description}
      \item{1.\ \ }The halo at high Galactic latitudes ($b = [-90\degr,-40\degr]$) 
      characterised by low emission fields with steep spectra (angular slope $\beta = [-3.0, -2.0]$),
      that is smooth emission dominated by large scale structures.  The slope is
      almost uniformly distributed within a wide range,  with median $\beta_{\rm med} = -2.6$.
      \item{2.\ \ }A transition region at mid-latitudes  ($b = [-40\degr,-20\degr]$)
      whose angular spectra are steep like those of the halo, but shows an 
      increase of the emission power with decreasing latitude. 
      \item{3.\ \ }The disc at low latitudes ($b = [-20\degr, 0\degr]$) characterized by inverted spectra
      with slopes in a narrow range of median $\beta_{\rm med} = -1.8$.
     The amplitudes are two orders of magnitude brighter than in the halo and the power 
     gradually increases towards the Galactic plane. 
\end{description}      

The change in the angular slope around $b=-20\degr$ is abrupt
and identifies a sharp disc-halo transition from the smooth emission of 
the mid-high latitudes to the more complex behaviour of the disc; this is likely related
to the more turbulent and complex structure of the ISM in the disc.

The halo section has no clear trend with latitude of
either emission power or angular slope, and, at least along the
meridian sampled by PGMS, can be considered a single environment.
This is very important for CMB investigations, as it indicates
that it is possible to find large areas with optimal conditions for seeking the $B$--Mode.

The synchrotron emission of the whole halo section is very weak.
Once scaled to 70~GHz it is equivalent to $r = 3.3\times10^{-3}$,
so that an experiment aiming for a detection limit of $\delta r=0.01$--0.02
would need no synchrotorn foreground cleaning.

The dust component is also faint and equal to
the synchrotron emission at 60--80~GHz for polarization fractions between 5 and 20~\%. 
The frequency of minimum foreground in this low emission region is thus similar to that 
found with WMAP for the general high Galactic latitudes (75\% of the sky).
If confirmed in other regions, this would imply both that the dust-synchrotron
power ratio is rather independent of the brightness of the Galactic emission and  the
frequency of minimum foreground  nearly independent of the sky position. 

We estimate the $r$ detection limit of this area accounting for
the use of foreground cleaning procedures. We apply both
a coarse and a more refined method. 
The Galactic emission is so low that the dominant residual
contamination is from gravitational
lensing, even assuming a 10-fold reduction of the lensing foreground from cleaning.
For both the two cleaning methods the detection limit is
$\delta r \sim 2\times 10^{-3}$ (3-sigma C.L.) if the CMB $B$--Mode search is
conducted at  70~GHz.
This result provides a sound basis for investigating the $B$--Mode. 
The detection limit we have found here is even better than the goals of 
the most advanced sub-orbital experiments ($r \sim 0.01$, e.g.: SPIDER, EBEX, 
and QUIET, \citealt{crill08,grainger08,samtleben08}) 
and proves that there exists at least one area of the sky where 
it is realistic to carry out investigations
of the $B$--mode down to very low limits of $r$.

At 150 GHz the detection limit rises, but is still 
better than $\delta r = 0.01$ assuming a reasonable 
dust polarization fraction ($f_{\rm pol} <$~12\%).  

These results are valid in the area actually observed by our survey.
However, the PGMS halo section is extended over $50\degr$ along one dimension and it is
unlikely that its excellent conditions are confined to its $5\degr$
width.
We have explored the likely results from a larger $50\degr \times
50\degr$ area, having properties similar to those of the PGMS halo section.
In such a region the $r$ detection limit would drop to
$\delta r = 5\times 10^{-4}$ (3-sigma) at 70~GHz.

It is worth noticing that the gravitational lensing needs to be cleaned to take advantage 
of the low Galactic emission of the PGMS halo section. This can be effectively carried out only with 
high resolution data ($10'$ or finer, \citealt{seljak04}) and the design of CMB experiments 
should comply with that rather than be limited to 1-degree to fit the peak at $\ell \sim 90$.

The results obtained here might suggest a review of plans to detecting 
the CMB $B$--Mode and associated investigations of the inflationary scenarios.
While the detection limit is limited to $\delta r = 1 \times 10^{-2}$
by the detector array size and sensitivity, observations at 150~GHz might be sufficient if
conducted in an area like the PGMS with clear advantages 
of using the currently best detectors (bolometers) and of an 
experiment  more compact than at 70~GHz. 
Moreover, the synchrotron emission is sufficiently
weak at 150~GHz not to require any cleaning, which removes the need for
 low frequency channels to measure it. 
Experiments like EBEX and BICEP already
match such conditions, not only because of the design choices, but also because their
target areas intersects the PGMS strip \citep{grainger08,chiang09}.

A deeper detection limit,  down to $\delta r = 2 \times 10^{-3},$  could be reached by 
a sub-orbital experiment observing the same area but with the {\it CMB channel} shifted to 
70~GHz. The inclusion of a channel at a lower frequency would be required to measure the synchrotron component.
Finally, detections limits down
to $\delta r = 5 \times 10^{-4}$ coiuld be achieved by observing 
at 70~GHz in a large area of $50\degr \times 50\degr$ having the PGMS foreground levels.  
The location of the most suitable region must be determined, a task that can be accomplished 
by the forthcoming large foreground surveys like the S-band Polarization All Sky Survey (S-PASS), 
or the C-band All Sky Survey (C-BASS).

These limits are comparable 
to the goals of the space missions currently under study
such as B-POL and CMBPol \citep{debernardis08,baumann08},
but with the significant advantage that 
such an area is still sufficiently compact to be observable by a
sub-orbital experiment.
The limit $r \sim 1 \times 10^{-3}$ is an important threshold for the inflationary physics
since it is about the lower limit of the important class of
inflationary models with low degree of fine tuning \citep{boyle06}.
Our study shows that this threshold may be reached with an 
easier and cheaper sub-orbital experiment rather than a more complex space mission,
making this goal more realistically achievable with a smaller budget and in a shorter time
than that required to develop space-borne equipment.

\vskip 0.7cm

The PGMS data will be made available at the site {\it http://www.atnf.csiro.au/people/Ettore.Carretti/PGMS}

\section*{Acknowledgments}

This work has been partly supported by the project SPOrt  
funded by the Italian Space Agency (ASI) and by the ASI contract I/016/07/0 {\it COFIS}.
M.H. acknowledges support from the National Radio Astronomy
Observatory (NRAO), which is operated by Associated Universities Inc.,
under cooperative agreement with the National Science Foundation.
We wish to thank Warwick Wilson for his support in the DFB1 set-up, 
John Reynolds for the observations set-up, and an anonymous referee 
for useful comments.
Part of this work is based on observations taken with
the Parkes Radio Telescope, which is part of the Australia Telescope,
funded by the Commonwealth of Australia for operation
as a National Facility managed by CSIRO.
We acknowledge the use of the CMBFAST
and HEALPix packages.

\bsp

\label{lastpage}


\begin{thebibliography}{}
\bibitem[\protect\citeauthoryear{Amarie, Hirata \& Seljak}{Amarie et al.}{2005}]{amarie05} 
    Amarie~M., Hirata~C., Seljak~U., 2005, PRD, 72, 123006 
\bibitem[\protect\citeauthoryear{Baumann et al.}{2008}]{baumann08}	
    Baumann D., et al., 2008, arXiv:0811.3911 [astro-ph]
\bibitem[\protect\citeauthoryear{Beck}{2008}]{beck08} 
    Beck~R., 2008, in "The UV Window to the Universe", eds. A.I.~Gomez de Castro and M.~Castellanos, 
    Ap\&SS, in press, arXiv:0711.4700 [astro-ph]
\bibitem[\protect\citeauthoryear{Beno$\hat{\rm\i}$t et al.}{2004}]{be04}	
    Beno$\hat{\rm\i}$t~A., et al., 2004, A\&A, 424, 571
\bibitem[\protect\citeauthoryear{Bernardi et al.}{2004}]{bernardi04}
    Bernardi~G., Carretti~E., Fabbri~R., Sbarra~C., Poppi~S., Cortiglioni~S., Jonas~J.L.,
     2004, MNRAS, 351, 436 
\bibitem[\protect\citeauthoryear{Bernardi et al.}{2006}]{bernardi06}
    Bernardi~G., Carretti~E., Sault~R.J., Cortiglioni~S., Poppi~S., 2006, MNRAS, 370, 2064
\bibitem[\protect\citeauthoryear{Boyle, Steinhardt, \& Turok}{Boyle et al.}{2006}]{boyle06}
    Boyle L.A., Steinhardt P.J., \& Turok N., 2006, Phys. Rev. Lett. 96, 111301 
\bibitem[\protect\citeauthoryear{Brown et al.}{2007}]{brown07} 
    Brown~J.C., Haverkorn~M., Gaensler~B.M., Taylor~A.R., Bizunok~N.S., McClure-Griffiths~N.M., 
    Dickey~J.M., Green~A.J., 2007, ApJ, 663, 258 
\bibitem[\protect\citeauthoryear{Brown et al.}{2009}]{brown09} 
    Brown~M.L., et al, 2009, ApJ, 705, 978 
\bibitem[\protect\citeauthoryear{Beuermann, Kanbach \& Berkhuijsen}{Beuermann et al.}{1985}]{beuermann85}
    Beuermann K., Kanbach G., \& Berkhuijsen E.M., 1985, A\&A, 153, 17
\bibitem[\protect\citeauthoryear{Carretti et al.}{2005a}]{carretti05a}
    Carretti E., Bernardi G., Sault R.J., Cortiglioni S., 
        \& Poppi S., 2005a, MNRAS, 358, 1 
\bibitem[\protect\citeauthoryear{Carretti et al.}{2005b}]{carretti05b}
    Carretti E., McConnell D., McClure-Griffiths N.M., Bernardi G., Cortiglioni S., 
        \& Poppi S., 2005b, MNRAS, 360, L10 
\bibitem[\protect\citeauthoryear{Carretti et al.}{2006a}]{carretti06a}
    Carretti E., Poppi S., Reich W., Reich P., F\"urst E., 
        Bernardi G., Cortiglioni S., Sbarra C., 2006a, MNRAS, 367, 132
\bibitem[\protect\citeauthoryear{Carretti, Bernardi \& Cortiglioni}{Carretti et al.}{2006b}]{carretti06b} 
    Carretti E., Bernardi G., Cortiglioni S., 2006b, MNRAS, 373, L93 
\bibitem[\protect\citeauthoryear{Chiang et al.}{2009}]{chiang09} 
    Chiang ~H.C., et al, 2009, submitted to ApJ, 	arXiv:0906.1181 [astro-ph.CO] 
\bibitem[\protect\citeauthoryear{Crill et al.}{2008}]{crill08} 
    Crill~B.P., et al., 2008, in Space Telescopes and Instrumentation 2008: Optical, Infrared, and Millimeter,
    Eds. J.M.~Oschmann~Jr., M.W.M.~de Graauw \& H.A.~MacEwens, Proc. of SPIE, 7010, 70102P
\bibitem[\protect\citeauthoryear{de Bernardis et al.}{2008}]{debernardis08} 
    de Bernardis~P., Bucher~M., Burigana~C., Piccirillo~L., 2008, Exp. Astron., in press, arXiv:0808.1881 [astro-ph]
\bibitem[\protect\citeauthoryear{Emerson \& Gr\"ave}{1988}]{em88} 
     Emerson D.T., Gr\"ave R., 1988, A\&A,190, 353
\bibitem[\protect\citeauthoryear{Finkbeiner, Davis \& Schlegel}{Finkbeiner et al.}{1999}]{fi99}
   Finkbeiner D.P., Davis M., Schlegel D.J., 1999, ApJ, 524, 867
\bibitem[\protect\citeauthoryear{Frick et al.}{2001}]{frick01}
   Frick P., Stepanov R., Shukurov A., \& Sokoloff D., 2001, MNRAS, 325, 649
\bibitem[\protect\citeauthoryear{Gaensler et al.}{2001}]{gaensler01}
  Gaensler B.M., Dickey J.M., McClure-Griffiths N.M., Green A.J., Wieringa M.H., \& Haynes R.F., 2001, ApJ, 549, 959
\bibitem[\protect\citeauthoryear{Gold et al.}{2009}]{gold09}
   Gold~B., et al., 2009, ApJS, in press, arXiv:0803.0715 [astro-ph]
\bibitem[\protect\citeauthoryear{G\'orski et al.}{2005}]{go05}
    G\'orski K.M., et al., 2005, ApJ, 622, 759
\bibitem[\protect\citeauthoryear{Grainger et al.}{2008}]{grainger08}
    Grainger~W., et al., 2008, in Millimeter and Submillimeter Detectors and Instrumentation for Astronomy IV,
    Eds. W.D.~Duncan, W.S.~Holland, S.~Withington, J.~Zmuidzinas, Proc. of SPIE Vol. 7020 70202N-2
\bibitem[\protect\citeauthoryear{Han}{2002}]{han02} 
    Han~J.L., 2002, in Astrophysical Polarized Backgrounds, eds. S.~Cecchini, S.~Cortiglioni, R.~J.~Sault, C.~Sbarra, 
    AIP Conf.  Ser., 609, 96
\bibitem[\protect\citeauthoryear{Han}{1997}]{han97}
  Han~J.L., Manchester~R.N., Berkhuijsen~E.M., \& Beck~R., 1997, A\&A, 322, 98
\bibitem[\protect\citeauthoryear{Han et al.}{2006}]{han06} 
    Han~J.L., Manchester~R.N., Lyne~A.G., Qiao~G.J., van Straten~W., 2006, ApJ, 642, 868
\bibitem[\protect\citeauthoryear{Haverkorn et al.}{2006}]{haverkorn06} 
    Haverkorn,~M., Gaensler~B.M., McClure-Griffiths~N.M., Dickey,~J.M., Green~A.J., 2006, ApJS, 167, 230 
\bibitem[\protect\citeauthoryear{Hinshaw et al.}{2009}]{hinshaw09} 
    Hinshaw~G., et al., 2009, ApJS, 180, 225
\bibitem[\protect\citeauthoryear{Hockney \& Eastwood}{1981}]{hockney81}
    Hockney~R.W., Eastwood~J.W., 1981, Computer Simulation Using Particles (New York: McGraw-Hill)
\bibitem[\protect\citeauthoryear{Jansson et al.}{2009}]{jansson09}
  Jansson~R., Farrar~G.R., Waelkens~A.H., \& Ensslin~T.A. 2009, JCAP, 7, 21
\bibitem[\protect\citeauthoryear{Kamionkowski \& Kosowsky}{1998}]{Kamionkowski98}
    Kamionkowski~M., Kosowsky~A., 1998, PRD, 57, 685 
\bibitem[\protect\citeauthoryear{Komatsu et al.}{2009}]{komatsu09} 
    Komatsu E., et al., 2009, ApJS, 180, 330
\bibitem[\protect\citeauthoryear{Kraus}{1986}]{kraus86} 
    Kraus~J.D., 1986. Radio Astronomy (Ed. CygnusÐQuasar Books: Powell, OH)
\bibitem[\protect\citeauthoryear{Lange}{2008}]{lange08}
    Lange A., 2008, in CMB component separation and the physics of foregrounds,
    http://planck.ipac.caltech.edu/content/ForegroundsConference/presentationsForWEB/58\_andrewLange.pdf
\bibitem[\protect\citeauthoryear{La Porta et al.}{2006}]{laporta06}
    La Porta~L., Burigana~C., Reich~W., Reich~P., 2006, A\&A, 455, L9
\bibitem[\protect\citeauthoryear{Masi et al.}{2006}]{masi06}
    Masi~S., et al., 2006, A\&A, 458, 687 
\bibitem[\protect\citeauthoryear{Page et al.}{2007}]{page07}
    Page~L., et al, 2007, ApJS, 170, 335
\bibitem[\protect\citeauthoryear{Peiris et al.}{2003}]{peiris03}
    Peiris~H.V., et al, 2003, ApJS, 148, 213
\bibitem[\protect\citeauthoryear{Ponthieu et al.}{2005}]{ponthieu05}
    Ponthieu N., et al., 2005, A\&A, 444, 327
\bibitem[\protect\citeauthoryear{Reynolds}{1994}]{reynolds94} 
    Reynolds~J.E. 1994, ATNF Tech. Doc. Ser. 39.3040
\bibitem[\protect\citeauthoryear{Samtleben}{2008}]{samtleben08} 
    Samtleben~D., in Cosmology 2008, Proc. of the 43rd Rencontres de Moriond, in press,
    arXiv:0806.4334 [astro-ph]
\bibitem[\protect\citeauthoryear{Sbarra et al.}{2003}]{sbarra03} 
    Sbarra~C., Carretti~E., Cortiglioni~S., Zannoni~M., Fabbri~R., 
    Macculi~C., Tucci~M., 2003, A\&A, 401, 1215
\bibitem[\protect\citeauthoryear{Seljak \& Hirata}{2004}]{seljak04} 
    Seljak~U., Hirata~C.M., 2004, PRD, 69, 043005 
\bibitem[\protect\citeauthoryear{Sun et al.}{2008}]{sun08} 
    Sun~X.H., Reich~W., Waelkens~A., En{\ss}lin~T.A., 2008, A\&A, 477, 573 
\bibitem[\protect\citeauthoryear{Tegmark}{1997}]{tegmark97a}
    Tegmark~M.,\ 1997, PRD, 56,  4514 
\bibitem[\protect\citeauthoryear{Tegmark, Taylor \&  Heavens}{Tegmark et al.}{1997}]{tegmark97b} 
    Tegmark~M., Taylor~A.N., Heavens~A.F., 1997, ApJ, 480,  22 
\bibitem[\protect\citeauthoryear{Testori, Reich, \& Reich}{Testori et al.}{2008}]{testori08} 
    Testori~J.C., Reich~P., Reich~W., 2008, A\&A, 484, 733 
\bibitem[\protect\citeauthoryear{Thomas et al.}{1997}]{thomas97}
   Thomas~B.M., Schafer~J.T., Sinclair~M.W., Kesteven~M.J., Hall~P.J., 1997,
    IEEE Antennas \& Propagation Magazine,  39, 54
\bibitem[\protect\citeauthoryear{Tucci et al.}{2005}]{tucci05} 
    Tucci~M., Mart{\'{\i}}nez-Gonz{\'a}lez~E., Vielva~P., Delabrouille~J., 2005, MNRAS, 360, 935 
\bibitem[\protect\citeauthoryear{Verde et al.}{2006}]{verde06}
    Verde~L., Peiris~H.V., Jimenez~R., 2006, JCAP, 1, 19
\bibitem[\protect\citeauthoryear{Wolleben et  al.}{2006}]{wolleben06} 
   Wolleben~M., Landecker~T.L., Reich~W., Wielebinski~R., 2006, A\&A, 448, 411 
\bibitem[\protect\citeauthoryear{Zaldarriaga}{1998}]{zaldarriaga98}
   Zaldarriaga M., 1998, Ph.D. Thesis, M.I.T., astro-ph/9806122
\end{thebibliography}
\end{document}